\documentclass[preprint]{aastex}
\usepackage{amsmath}
\usepackage{graphicx}
\usepackage{booktabs,threeparttable}
\usepackage{hyperref}
\usepackage{multirow}
\usepackage{ulem}
\usepackage{lscape}
\usepackage{textpos}
\shorttitle{Formation of Magnetized Prestellar Cores}
\shortauthors{Chen \& Ostriker}
\begin{document}
\title{Formation of Magnetized Prestellar Cores with Ambipolar Diffusion and Turbulence}
\author{Che-Yu Chen\altaffilmark{1} and Eve C. Ostriker\altaffilmark{1,2}}
\altaffiltext{1}{Department of Astronomy, University of Maryland, College Park, MD 20742}
\altaffiltext{2}{Department of Astrophysical Sciences, Princeton University, Princeton, NJ, 08544}
\email{cychen@astro.umd.edu, eco@astro.princeton.edu}

\begin{abstract}

We investigate the roles of magnetic fields and ambipolar diffusion during prestellar core formation in turbulent giant molecular clouds (GMCs), using three-dimensional numerical simulations. Our simulations focus on the shocked layer produced by a converging large-scale flow, and survey varying ionization and angle between the upstream flow and magnetic field. We also include ideal magnetohydrodynamic (MHD) and hydrodynamic models. From our simulations, we identify hundreds of self-gravitating cores that form within $1$~Myr, with masses $M\sim 0.04-2.5$~M$_\odot$ and sizes $L\sim 0.015-0.07$~pc, consistent with observations of the peak of the core mass function (CMF). Median values are $M=0.47~\mathrm{M}_\odot$ and $L=0.03$~pc. Core masses and sizes do not depend on either the ionization or upstream magnetic field direction. In contrast, the mass-to-flux ratio does increase with lower ionization, from twice to four times the critical value. The higher mass-to-flux ratio for low ionization is the result of enhanced transient ambipolar diffusion when the shocked layer first forms. However, ambipolar diffusion is not necessary to form low-mass supercritical cores. For ideal MHD, we find similar masses to other cases. These masses are $1-2$ orders of magnitude lower than the value $M_\mathrm{mag,sph}=0.007~B^3/ (G^{3/2}\rho^2)$ that defines a magnetically supercritical sphere under post-shock ambient conditions. This discrepancy is the result of anisotropic contraction along field lines, which is clearly evident in both ideal MHD and diffusive simulations. We interpret our numerical findings using a simple scaling argument which suggests that gravitationally critical core masses will depend on the sound speed and mean turbulent pressure in a cloud, regardless of magnetic effects.

\end{abstract}

\section{Introduction}
\label{sec:intro}

The formation of stars begins with dense molecular cores \citep{2007ARA&A..45..565M,2009sfa..book..254A}. These cores form through the concentration of overdense regions within turbulent, filamentary GMCs; subsequent core collapse leads to protostellar (or protobinary)/disk systems. Magnetic fields are important at all scales during this process \citep{2007ARA&A..45..565M,2012ARA&A..50...29C}: the cloud-scale magnetic field can limit compression in interstellar shocks that create dense clumps and filaments in which cores form, while the local magnetic field within individual cores can prevent collapse if it is large enough \citep{1956MNRAS.116..503M,1966MNRAS.132..359S,1976ApJ...210..326M}, and can help to remove angular momentum during the disk formation process if cores are successful in collapsing \citep{1985prpl.conf..320M,1991ApJ...373..169M, 2003ApJ...599..363A, 2013PPVI...Li}. The significance of magnetic fields in self-gravitating cores can be quantified by the ratio of mass to magnetic flux; only if the mass-to-flux ratio exceeds a critical value is gravitational collapse possible. How the mass-to-flux ratio increases from the strongly-magnetized interstellar medium to weakly-magnetized stars is a fundamental problem of star formation \citep{1987ARA&A..25...23S, 2007ARA&A..45..565M}. Here, as suggested in \citet[][hereafter CO12]{2012ApJ...744..124C}, we consider core formation in GMCs with highly supersonic turbulence and non-ideal MHD.

Magnetic fields are coupled only to charged particles, while the gas in GMCs and their substructures is mostly neutral. The ability of magnetic fields to affect core and star formation thus depends on the collisional coupling between neutrals and ions. Ambipolar diffusion is the non-ideal MHD process that allows charged particles to drift relative to the neutrals, with a drag force proportional to the collision rate \citep{1956pfig.book.....S}. Ambipolar drift modifies the dynamical effect of magnetic fields on the gas, and may play a key role in the star formation. 

In classical theory, quasi-static ambipolar diffusion is the main mechanism for prestellar cores to lose magnetic support and reach supercritical mass-to-flux ratios. Through ambipolar drift, the mass within dense cores can be redistributed, with the neutrals diffusing inward while the magnetic field threading the outer region is left behind \citep{1979ApJ...228..475M}. However, the quasi-static evolution model \citep[e.g.][]{1999osps.conf..305M,2001ApJ...547..272C} gives a prestellar core lifetime considerably longer (up to a factor of 10) than the gravitational free-fall timescale, $t_\mathrm{ff}$, while several observational studies have shown that cores only live for $(2-5)$ $t_\mathrm{ff}$ \citep[e.g.][]{2007prpl.conf...33W, 2009ApJS..181..321E}. 

The failure of the traditional picture to predict core lifetimes indicates that supercritical cores may not have formed quasi-statically through ambipolar diffusion. Indeed, it is now generally recognized that, due to pervasive supersonic flows in GMCs, core formation is not likely to be quasi-static. Realistic star formation models should take both ambipolar diffusion and large-scale supersonic turbulence into consideration. This turbulence may accelerate the ambipolar diffusion process \citep{2004ApJ...603..165H,2004ApJ...609L..83L}, with an analytic estimate of the enhanced diffusion rate by a factor of 2$-$3 for typical conditions in GMCs \citep{2002ApJ...570..210F}. 

In our previous work (\hyperlink{CO12}{CO12}), we investigated the physical mechanism driving enhanced ambipolar diffusion in one-dimensional C-type shocks. These shocks pervade GMCs, and are responsible for the initial compression of gas above ambient densities. We obtained a formula for the C-shock thickness as a function of density, magnetic field, shock velocity, and ionization fraction, and explored the dependence of shock-enhanced ambipolar diffusion on environment through a parameter study. Most importantly, we identified and characterized a transient stage of rapid ambipolar diffusion at the onset of shock compression, for one-dimensional converging flows. For an interval comparable to the neutral-ion collision time and before the neutral-ion drift reaches equilibrium, the neutrals do not experience drag forces from the ions. As a consequence, the initial shock in the neutrals is essentially unmagnetized, and the neutrals can be very strongly compressed. This transient stage, with timescale $t_\mathrm{transient}\sim 1$~Myr (but depending on ionization), can create dense structures with much higher $\rho/B$ than upstream gas. \hyperlink{CO12}{CO12} suggested this could help enable supercritical core formation.
\hyperlink{CO12}{CO12} also found that (1) the perpendicular component of the magnetic field is the main determinant of the shock compression, and (2) the perpendicular component of the magnetic field $B_\perp$ must be weak ($\lesssim 5~\mu$G) for transient ambipolar diffusion in shocks to significantly enhance $\rho/B_\perp$.

Observations of nearby clouds provide direct constraints on the role of magnetic fields, as well as other properties of prestellar cores. The typical mean mass-to-flux ratio of dark cloud cores is $\Gamma\sim 2$ (in units of critical value; see Equation~(\ref{GammaDef})) from Zeeman studies \citep{2008A&A...487..247F,2008ApJ...680..457T}. Due to the instrumental limitations, magnetic field observations in solar-mass and smaller scale regions are relatively lacking compared with observations of larger scales \citep[see review in][]{2012ARA&A..50...29C}, however. Surveys in nearby clouds have found that prestellar cores have masses between $\sim 0.1-10$~M$_\odot$ and sizes $\sim 0.01-1$~pc \citep[e.g.][]{2001A&A...372L..41M, 2009ApJ...691.1560I, 2009ApJ...699..742R,2013MNRAS.432.1424K}. In addition, a mass-size relation has been proposed as a power law $M\propto R^k$, with $k=1.2-2.4$ dependent on various molecule tracers \citep[e.g.][]{1996ApJ...471..816E,2010MNRAS.402..603C,2010ApJ...723..492R,2013MNRAS.432.1424K}.

The magnetic field strength within prestellar cores is important for late evolution during core collapse, since disk formation may be suppressed by magnetic braking (for recent simulations see \citealt{2003ApJ...599..363A,2008A&A...477....9H,2008ApJ...681.1356M,2011A&A...528A..72H}; or see review in \citealt{2013PPVI...Li}). However, many circumstellar disks and planetary systems have been detected \citep[e.g.][]{2001ApJ...553L.153H,2010A&A...512A..40M}, suggesting that the magnetic braking ``catastrophe" seen in many simulations does not occur in nature. The proposed solutions include the misalignment between the magnetic and rotation axes \citep[e.g.][]{2009A&A...506L..29H,2010MNRAS.409L..39C,2012A&A...543A.128J,2013ApJ...767L..11K}, turbulent reconnection and other turbulent processes during the rotating collapse \citep[e.g.][]{2012ApJ...747...21S,2012MNRAS.423L..40S,2013MNRAS.432.3320S}, and non-ideal MHD effects including ambipolar diffusion, Hall effect, and Ohmic dissipation \citep[e.g.][]{2010ApJ...716.1541K,2011ApJ...738..180L,2011PASJ...63..555M,2012A&A...541A..35D,2013ApJ...763....6T}. If prestellar cores have sufficiently weak magnetic fields, however, braking would not be a problem during disk formation \citep[e.g.][]{2008ApJ...681.1356M,2013PPVI...Li,2013ApJ...774...82L}. Therefore, the magnetic field (and mass-to-flux ratio) within a prestellar core is important not just for the ability of the core to collapse, but also of a disk to form.

Fragmentation of sheetlike magnetized clouds induced by small-amplitude perturbation and regulated by ambipolar diffusion has been widely studied \citep[e.g.][]{2000ApJ...532..361I,2004ApJ...607L..39B,2005ApJ...622..393B,2006ApJ...652..442C,2009NewA...14..221B}. Analogous fully three-dimensional simulations have also been conducted \citep[e.g.][]{2007MNRAS.380..499K}. Supercritical cores formed in the flattened layer have masses $\sim 0.1-10$~M$_\odot$ \citep[e.g.][]{2000ApJ...532..361I,2009NewA...14..221B}, at timescales $\sim 1-10$~Myr dependent on the initial mass-to-flux ratio of the cloud \citep[e.g.][]{2000ApJ...532..361I,2007MNRAS.380..499K,2009NewA...14..221B}. The above cited simulations start from relatively high densities ($\sim 10^4$~cm$^{-3}$; e.g. \citealt{2007MNRAS.380..499K}) and included only the low-amplitude perturbations. Alternatively, \citet{2004ApJ...609L..83L} and \citet{2005ApJ...631..411N} took the formation of these overdense regions into consideration by including a direct treatment of the large-scale supersonic turbulence. They demonstrated that ambipolar diffusion can be sped up locally by the supersonic turbulence, forming cores with masses $\sim 0.5$~M$_\odot$ and sizes $\sim 0.1$~pc within $\sim 2$~Myr, while the strong magnetic field keeps the star formation efficiency low ($1-10\%$). Similarly, \citet{2009NewA...14..483B} found that turbulence-accelerated, magnetically-regulated core formation timescales are $\sim 1$~Myr in two-dimensional simulations of magnetized sheet-like clouds, with corresponding three-dimensional simulations showing comparable results \citep{2008ApJ...679L..97K,2011ApJ...728..123K}. In addition,
\citet{2008ApJ...687..354N} measured the core properties in their three-dimensional simulations to find $L_\mathrm{core}\sim 0.04-0.14$~pc, $\Gamma_\mathrm{core}\sim 0.3-1.5$, and $M_\mathrm{core}\sim 0.15-12.5$~M$_\odot$, while \citet{2009NewA...14..483B} found a broader core mass distribution $M_\mathrm{core}\sim 0.04-25$~M$_\odot$ in their parameter study using thin-sheet approximation. 

Supersonic turbulence within GMCs extends over a wide range of spatial scales \citep{2004RvMP...76..125M,2007prpl.conf...63B}. Although turbulence contains sheared, diverging, and converging regions in all combinations, regions in which there is a large-scale convergence in the velocity field will strongly compress gas, creating favorable conditions for the birth of prestellar cores. \citet[][hereafter GO11]{2011ApJ...729..120G} investigated core formation in an idealized model containing both a large-scale converging flow and multi-scale turbulence. These simulations showed that the time until the first core collapses depends on inflow Mach number ${\cal M}$ as $t_\mathrm{collapse}\propto {\cal M}^{-1/2}$. With a parameter range ${\cal M} = 1.1$ to $9$,
cores formed in the \hyperlink{GO11}{GO11} simulations had masses $0.05-50$~M$_\odot$.
Following similar velocity power spectrum but including ideal MHD effects, \citet{2014arXiv1401.6096M} performed simulations with sink particle, radiative transfer, and protostellar outflows to follow the protostar formation in turbulent massive clump. They demonstrated that the median stellar mass in the simulated star cluster can be doubled by the magnetic field, from $0.05$~M$_\odot$ (unmagnetized case) to $0.12$~M$_\odot$ (star cluster with initial mass-to-flux ratio $\Gamma=2$).
This is qualitatively consistent with the conclusion in \citet{2013ApJ...774L..31I}, that the mass of the cores formed in the post-shock regions created by cloud-cloud collision is positively related to (and dominated by) the strong magnetic field in the shocked layer. Note that, though the main focus of \citet{2013ApJ...774L..31I} is the cloud's ability to form massive cores ($\sim 20-200$~M$_\odot$ in their simulations), the idea of cloud-cloud collision is very similar to the converging flows setup adopted in \hyperlink{GO11}{GO11} and this study.

In this paper, we combine the methods of \hyperlink{CO12}{CO12} for modeling ambipolar diffusion with the methods of \hyperlink{GO11}{GO11} for studying self-gravitating structure formation in turbulent converging flows. Our numerical parameter study focuses on the level of ambipolar diffusion (controlled by the ionization fraction of the cloud) and the obliquity of the shock (controlled by the angle between the magnetic field and the upstream flow). We show that filamentary structures similar to those seen in observations \citep[see review in][]{2013PPVI...Andre} develop within shocked gas layers, and that cores form within these filaments. We measure core properties to test their dependence on these parameters. As we shall show, our models demonstrate that low-mass supercritical cores can form for all magnetic obliquities and all levels of ionization, including ideal MHD. However, our models also show that ambipolar diffusion affects the magnetization of dynamically-formed cores.

The outline of this paper is as follows. We provide a theoretical analysis of oblique MHD shocks in Section~\ref{theory}, pointing out that a quasi-hydrodynamic compression ratio (which is $\sim 5$ times stronger than in \textit{fast} MHD shocks for the parameters we study) can exist when the converging flow is nearly parallel to the magnetic field. We also show that shock compression cannot increase the mass-to-flux ratio except in the nearly-parallel case or with ambipolar diffusion. Section~\ref{sec: methods} describes methods used in our numerical simulations and data analysis, including our model parameter set and method for measuring magnetic flux within cores. The evolution of gas structure (including development of filaments) and magnetic fields for varying parameters is compared in Section~\ref{sec::evolution}. In Section~\ref{sec:results} we provide quantitative results for masses, sizes, magnetizations, and other physical properties of the bound cores identified from our simulations. Implications of these results for core formation is discussed in Section~\ref{sec::CF}, where we argue that the similarity of core masses and sizes among models with different magnetizations and ionizations can be explained by anisotropic condensation preferentially along the magnetic field. Section~\ref{sec: summary} summarizes our conclusions.

\section{Theoretical Analysis}
\label{theory}

\subsection{Oblique MHD Shock}
\label{sec: obshock}

\begin{figure}
\begin{center}
\includegraphics[scale=0.6]{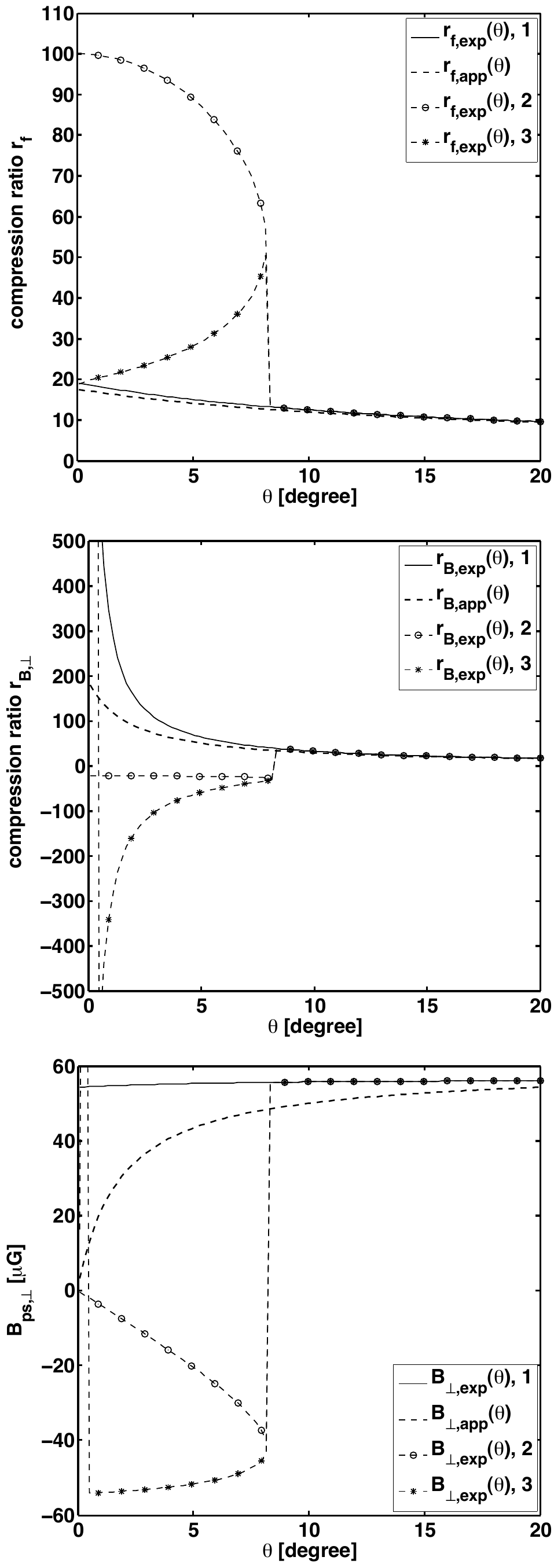}
\caption{Multiple solutions for Equation~(\ref{rfexp}) at varying $\cos\theta=\hat{\mathbf{B}}\cdot\hat{\mathbf{v}}$ with the following parameters: ${\cal M} = 10$, $B_0 = 10$~$\mu$G, $\rho_0 = \mu_n\cdot 1000$~cm$^{-3}$ where $\mu_n = 2.3$~$m_\mathrm{H}$. \textit{Top:} Compression ratio for neutrals. Equation~(\ref{rfappO1}) works as a good analytical approximation to $r_{f, \mathrm{exp}}(\theta), 1$. \textit{Middle:} Compression ratio for the perpendicular component (with respect to the inflow direction) of the magnetic field. The analytical approximation $r_{B, \mathrm{app}}(\theta)$ is calculated from Equation~(\ref{rBf}), using Equation~(\ref{rfappO1}) for $r_f(\theta)$. \textit{Bottom}: The corresponding post-shock magnetic field component that is perpendicular to the inflow (parallel to the shock front).}
\label{rfcomp}
\end{center}
\end{figure}

\hyperlink{CO12}{CO12} describe a one-dimensional simplified MHD shock system with velocity and magnetic field perpendicular to each other, including a short discussion of oblique shocks. Here we review the oblique shock equations and write them in a more general form to give detailed jump conditions.

We shall consider a plane-parallel shock with uniform pre-shock neutral density $\rho_0$ and ionization-recombination equilibrium everywhere. The shock front is in the $x$-$y$ plane, the upstream flow is along the $z$-direction ($\mathbf{v}_{0} = v_{0}\hat{\mathbf{z}}$), and the upstream magnetic field is in the $x$-$z$ plane, at an angle $\theta$ to the inflow ($\mathbf{B}_0 = B_0\sin\theta \hat{\mathbf{x}} + B_0\cos\theta \hat{\mathbf{z}}$) such that $B_x=B_0\sin\theta$ is the upstream component perpendicular to the flow. The parameters ${\cal M}$ and $\beta$ (upstream value of the Mach number and plasma parameter) defined in \hyperlink{CO12}{CO12} therefore become 
\begin{equation}
{\cal M} \equiv {\cal M}_z = \frac{v_{0}}{c_{s}},\ \ \ \frac{1}{\beta_0} \equiv  \frac{B_{0}^2}{8\pi\rho_{0}c_{s}^2} = \frac{1}{\beta_x} \frac{1}{\sin^2\theta}.
\label{define}
\end{equation}
The jump conditions of MHD shocks are described by compression ratios of density and magnetic field: 
\begin{equation}
r_f \equiv \frac{\rho_{n,\ \mathrm{downstream}}}{\rho_{n,\ \mathrm{upstream}}}, \ \ \ r_{B_\perp} \equiv \frac{B_{\perp,\ \mathrm{downstream}}}{B_{\perp,\ \mathrm{upstream}}}.
\end{equation}

From Equations~(A10) and (A14) in \hyperlink{CO12}{CO12}, we have
\begin{equation}
\frac{\sin^2\theta {r_f}^2}{\beta_0}\left(1-\frac{2\cos^2\theta}{\beta_0{\cal M}^2}\right)^2 = \left({\cal M}^2 + 1 + \frac{\sin^2\theta}{\beta_0} - \frac{{\cal M}^2}{r_f} - r_f\right)\left(1-\frac{2 r_f \cos^2\theta}{\beta_0{\cal M}^2}\right)^2,\label{rfexp}
\end{equation}
which can be solved numerically to obtain explicit solution(s) $r_{f, \mathrm{exp}}(\theta)$. The compression ratio for the magnetic field perpendicular to the inflow is
\begin{equation}
r_{B_\perp}(\theta) = r_f(\theta)\frac{1 - \frac{2\cos^2\theta}{\beta_0{\cal M}^2}}{1 - \frac{2 r_f(\theta) \cos^2\theta}{\beta_0{\cal M}^2}}.
\label{rBf}
\end{equation}
Equation~(A17) of \hyperlink{CO12}{CO12} gives an analytical approximation to $r_f(\theta)$:
\begin{equation}
r_{f, \mathrm{app}}(\theta) = \frac{\sqrt{\beta_0}{\cal M}}{\sin\theta}\left[\frac{2\sin\theta}{\sqrt{\beta_0}{\cal M}\tan^2\theta} + \frac{\sqrt{\beta_0}}{2{\cal M}\sin\theta} + 1\right]^{-1}. \label{rfappO1}
\end{equation}
Since Equation~(\ref{rfexp}) is a quartic function of $\theta$, there are four possible roots of $r_f$ for each angle, and $r_f(\theta)=const.=1$ (no-shock solution) is always a solution. When $\theta$ is large, Equation~(\ref{rfexp}) has one simple root ($r_f=1$) and a multiple root with multiplicity $=3$. When $\theta$ drops below a critical value, $\theta_\mathrm{crit}$, Equation~(\ref{rfexp}) has four simple roots, which give us four different values of $r_{B_\perp}$. Figure~\ref{rfcomp} shows the three explicit solutions for $r_f$ and $r_{B_\perp}$ ($r_{f,\mathrm{exp}}(\theta)$ and $r_{B,\mathrm{exp}}(\theta)$) as well as the approximations ($r_{f,\mathrm{app}}(\theta)$ and $r_{B,\mathrm{app}}(\theta)$) that employ Equation~(\ref{rfappO1}). 

The fact that there are multiple solutions for post-shock properties is the consequence of the non-unique Riemann problem in ideal MHD \citep[see discussions in e.g.][]{2003JPlPh..69..253T,2011JPlPh..77..207D,2013JPlPh..79..335T}, and whether all solutions are physically real is still controversial. The first set of solutions $r_{f, \mathrm{exp}}(\theta), 1$ and $r_{B, \mathrm{exp}}(\theta), 1$ shown in Figure~\ref{rfcomp} gives positive $r_f$ and $r_{B_\perp}$, classified as \textit{fast} MHD shocks \citep{1992phas.book.....S,1993ARA&A..31..373D}, and is the principal oblique shock solution referred to in this contribution\footnote{We use Equation~(\ref{rfappO1}) as analytical approximation for $r_f(\theta)$, if necessary.}. The other two solutions for post-shock magnetic field, $r_{B, \mathrm{exp}}(\theta), 2$ and $r_{B, \mathrm{exp}}(\theta), 3$, both become negative when $\theta < \theta_\mathrm{crit}$, indicating that the tangential component of the magnetic field to the shock plane is reversed in the post-shock region.
These two solutions are commonly specified as \textit{intermediate} shocks \citep[e.g.][]{1987GeoRL..14..668W,1995AdSpR..15..507K,2007PThPh.118...47I}.
Among these two field-reversal solutions, we notice that $r_{f, \mathrm{exp}}(\theta), 2$ approaches the hydrodynamic jump condition ($r_{f,\mathrm{hydro}} = {\cal M}^2$) when $\theta\rightarrow 0$, and $r_{B, \mathrm{exp}}(\theta), 2$ is smaller in magnitude than other solutions when $\theta < \theta_\mathrm{crit}$. Thus, we classify this set of solutions $r_{f, \mathrm{exp}}(\theta), 2$ and $r_{B, \mathrm{exp}}(\theta), 2$ as the \textit{quasi-hydrodynamic} shock. This quasi-hydrodynamic solution can create gas compression much stronger than the regularly-applied \textit{fast} shock condition, and may be the reason that when $\theta < \theta_\mathrm{crit}$, even ideal MHD simulations can generate shocked layers with relatively high mass-to-flux ratio (see Sections~\ref{sec::evolution} and \ref{sec:results} for more details).

The definition of $\theta_\mathrm{crit}$ can be derived from Equation~(\ref{rBf}), which turns negative when $1 - \frac{2\cos^2\theta}{\beta_0{\cal M}^2} > 0$ and $1 - \frac{2 r_f(\theta) \cos^2\theta}{\beta_0{\cal M}^2} <0$: 
\begin{equation}
\cos^2\theta > \cos^2\theta_\mathrm{crit} = \frac{\beta_0{\cal M}^2}{2 r_f(\theta_\mathrm{crit})}.
\end{equation}
Using Equation~(\ref{rfappO1}) and considering only the terms $\sim {\cal M}$, this becomes
\begin{equation}
\frac{\cos^2\theta_\mathrm{crit}}{\sin\theta_\mathrm{crit}} \approx \frac{\sqrt{\beta_0} {\cal M}}{2},
\end{equation}
or
\begin{equation}
\sin^2\theta_\mathrm{crit} + \frac{\sqrt{\beta_0} {\cal M}}{2} \sin\theta_\mathrm{crit} -1 = 0.
\label{thetaCrit}
\end{equation}
Assuming $\theta_\mathrm{crit}\ll 1$, this gives
\begin{equation}
\theta_\mathrm{crit} \sim \frac{2}{\sqrt{\beta_0} {\cal M}} = \sqrt{2}\frac{v_\mathrm{A,0}}{v_0},
\label{thetaC}
\end{equation}
where $v_\mathrm{A,0} \equiv B_0/\sqrt{4\pi\rho_0}$ is the Alfv{\'e}n speed in the cloud. Therefore, the criterion to have multiple solutions, $\theta < \theta_\mathrm{crit}$, is approximately equivalent to
\begin{equation}
v_\perp = v_0\sin\theta \lesssim v_0 \cdot \sqrt{2}\frac{v_\mathrm{A,0}}{v_0} \sim v_\mathrm{A,0}
\label{vperp}
\end{equation}
where $v_\perp$ is the component of the inflow perpendicular to the magnetic field. Though Equation~(\ref{thetaC}) only provides a qualitative approximation\footnote{For parameters used in Figure~\ref{rfcomp}, Equation~(\ref{thetaC}) gives $\theta_\mathrm{crit} = 18^\circ$, approximately $2$ times larger than the exact solution.} for $\theta_\mathrm{crit}$, Equation~(\ref{vperp}) suggests that when $v_\perp/v_\mathrm{A,0}$ is sufficiently small, high-compression {\it quasi-hydrodynamic} shocks are possible.

\subsection{Gravitational Critical Scales in Spherical Symmetry}
\label{spherical}

For a core to collapse gravitationally, its self-gravity must overcome both the thermal and magnetic energy. For a given ambient density $\rho\equiv\mu_n n$ and assuming spherical symmetry, the mass necessary for gravity to exceed the thermal pressure support (with edge pressure $\rho {c_s}^2$) is the mass of the critical Bonnor-Ebert sphere \citep[see e.g.][]{2009ApJ...699..230G}:
\begin{equation}
M_\mathrm{th,sph} = 4.18\frac{{c_s}^3}{\sqrt{4\pi G^3 \rho}} = 4.4~\mathrm{M}_\odot\left(\frac{T}{10~\mathrm{K}}\right)^{3/2}\left(\frac{n}{1000~\mathrm{cm}^{-3}}\right)^{-1/2}
\label{MthSPH}
\end{equation}
(see Section~\ref{sec::parameters} for discussion about the value of $\mu_n$). The corresponding length scale at the original ambient density is
\begin{equation}
R_\mathrm{th,sph} \equiv \left(\frac{3 M_\mathrm{th,sph}}{4\pi\rho}\right)^{1/3} = 2.3 \frac{c_s}{\sqrt{4\pi G \rho}} = 0.26~\mathrm{pc} \left(\frac{T}{10~\mathrm{K}}\right)^{1/2}\left(\frac{n}{1000~\mathrm{cm}^{-3}}\right)^{-1/2},
\label{RthSPH}
\end{equation}
although the radius of a Bonnor-Ebert sphere with mass given by Equation~(\ref{MthSPH}) would be smaller than Equation~(\ref{RthSPH}) by $25\%$, due to internal stratification.

In a magnetized medium with magnetic field $B$, the ratio of mass to magnetic flux for a region to be magnetically supercritical\footnote{See Section~\ref{sec: MagFluxCal} for more detailed discussion about the critical value of $M/\Phi_B$.} can be written as 
\begin{equation}
\frac{M}{\Phi_B}\bigg|_\mathrm{mag,crit} \equiv \frac{1}{2\pi\sqrt{G}}.
\label{GammaSPH}
\end{equation}
With $M=4\pi R^3 \rho /3$ and $\Phi_B = \pi R^2 B$ for a spherical volume at ambient density $\rho$, this gives 
\begin{equation}
M_\mathrm{mag,sph} = \frac{9}{128\pi^2G^{3/2}}\frac{B^3}{\rho^2} = 14~\mathrm{M}_\odot \left(\frac{B}{10~\mu\mathrm{G}}\right)^3\left(\frac{n}{1000~\mathrm{cm}^{-3}}\right)^{-2}.
\label{McritSPH}
\end{equation}
and
\begin{equation}
R_\mathrm{mag,sph} = \frac{3}{8\pi\sqrt{G}}\frac{B}{\rho} = 0.4~\mathrm{pc} \left(\frac{B}{10~\mu\mathrm{G}}\right) \left(\frac{n}{1000~\mathrm{cm}^{-3}}\right)^{-1},
\label{RcritSPH}
\end{equation}
A spherical region must have $M>M_\mathrm{th,sph}$ as well as $M>M_\mathrm{mag, sph}$ to be able to collapse. In the cloud environment (the pre-shock region), $B\sim 10~\mu$G and $n\sim 1000$~cm$^{-3}$ are typical. Comparing Equation~(\ref{MthSPH}) and (\ref{McritSPH}), the magnetic condition is more strict than the thermal condition; if cores formed from a spherical volume, the mass would have to exceed $\sim 10$~M$_\odot$ in order to collapse. This value is much larger than the typical core mass ($\sim 1$~M$_\odot$) identified in observations. This discrepancy is the reason why traditionally ambipolar diffusion is invoked to explain how low-mass cores become supercritical.

We can examine the ability for magnetically supercritical cores to form isotropically in a post-shock layer. The normalized mass-to-flux ratio
\begin{equation}
\Gamma\equiv \frac{M}{\Phi_B}\cdot 2\pi\sqrt{G}
\label{GammaDef}
\end{equation}
of a spherical volume with density $\rho$, magnetic field $B$, and mass $M$ is
\begin{align}
\Gamma_\mathrm{sph} &=\frac{8\pi\sqrt{G}}{3}\left(\frac{3}{4\pi}\right)^{1/3} M^{1/3}\rho^{2/3}B^{-1}\notag\\
&= 0.4 \left(\frac{M}{\mathrm{M}_\odot}\right)^{1/3}\left(\frac{n}{1000~\mathrm{cm}^{-3}}\right)^{2/3}\left(\frac{B}{10~\mu\mathrm{G}}\right)^{-1}.
\end{align}
Or, with $\Sigma = 4R\rho/3\equiv \mu_n N_n$ for a sphere, we have
\begin{equation}
\Gamma_\mathrm{sph} =2\pi\sqrt{G}\cdot\frac{\Sigma}{B} = 0.6\left(\frac{N_n}{10^{21}~\mathrm{cm}^{-2}}\right)\left(\frac{B}{10~\mu\mathrm{G}}\right)^{-1}.
\end{equation}
Considering the cloud parameters from Figure~\ref{rfcomp} (${\cal M} = 10$, $B_0 = 10$~$\mu$G, $n_0 = 1000$~cm$^{-3}$), the post-shock density and magnetic field are approximately $n_\mathrm{ps}\sim 10^4$~cm$^{-3}$ and $B_\mathrm{ps}\sim 50$~$\mu$G when $\theta>\theta_\mathrm{crit}$. A solar-mass spherical region in this shocked layer will have $\Gamma_\mathrm{ps,sph}\approx 0.37$; spherical contraction induced by gravity would be suppressed by magnetic fields. Thus, typical post-shock conditions are unfavorable for forming low-mass cores by spherical contraction in ideal MHD.

Furthermore, using $r_f$ and $r_{B_\perp}$ defined in Section~\ref{sec: obshock}, we can compare $\Gamma_\mathrm{ps,sph}$ and the pre-shock value $\Gamma_\mathrm{pre,sph}$ for spherical post-shock and pre-shock regions:
\begin{equation}
\frac{\Gamma_\mathrm{ps,sph}}{\Gamma_\mathrm{pre,sph}} = \left(\frac{M_\mathrm{ps}}{M_\mathrm{pre}}\right)^{1/3}\left(\frac{\rho_\mathrm{ps}}{\rho_\mathrm{pre}}\right)^{2/3}\left(\frac{B_\mathrm{ps}}{B_\mathrm{pre}}\right)^{-1} \approx \left(\frac{M_\mathrm{ps}}{M_\mathrm{pre}}\right)^{1/3} {r_f}^{2/3} {r_{B_\perp}}^{-1}.
\label{GammaComp}
\end{equation}
Considering volumes containing similar mass, $M_\mathrm{ps}\sim M_\mathrm{pre}$, the ratio between the post-shock and pre-shock $\Gamma_\mathrm{sph}$ is smaller than unity when $\theta > \theta_\mathrm{crit}$, because Equation~(\ref{rBf}) shows that $r_{B_\perp}$ is larger than $r_f$. Thus, provided $\theta > \theta_\mathrm{crit}$, the post-shock layer will actually have stronger magnetic support than the pre-shock region for a given spherical mass.

Based on the above considerations, formation of low-mass supercritical cores appears difficult in ideal MHD. Adapting classical ideas, one might imagine that low-mass subcritical cores form quasi-statically within the post-shock layer, then gradually lose magnetic support via ambipolar diffusion to become magnetically supercritical in a timescale $\sim 1-10~$Myr. A process of this kind would, however, give prestellar core lifetimes longer than observed, and most cores would have $\Gamma < 1$ (inconsistent with observations).

Two alternative scenarios could lead to supercritical core formation in a turbulent magnetized medium. First, the dynamic effects during a turbulence-induced shock (including rapid, transient ambipolar diffusion and the quasi-hydrodynamic compression when $\theta<\theta_\mathrm{crit}$) may increase the compression ratio of neutrals, creating $r_f \gg r_{B_\perp}$ and $\Gamma_\mathrm{ps,sph} > 1$, enabling low-mass supercritical cores to form. Second, even if the post-shock region is strongly magnetized, mass can accumulate through anisotropic condensation along the magnetic field until both the thermal and magnetic criteria are simultaneously satisfied. In this study, we carefully investigate these two scenarios, showing that both effects contribute to the formation of low-mass supercritical cores within timescale $\lesssim 0.6$~Myr, regardless of ionization or magnetic obliquity. 

\section{Numerical Methods and Models}
\label{sec: methods}

\subsection{Simulation Setup and Equations}

To examine core formation in shocked layers of partially-ionized gas, we employ a three-dimensional convergent flow model with ambipolar diffusion, self-gravity, and a perturbed turbulent velocity field. We conducted our numerical simulations using the \textit{Athena} MHD code \citep{2008ApJS..178..137S} with Roe's Riemann solver. To avoid negative densities if the second-order solution fails, we instead use first-order fluxes for bad zones. The self-gravity of the domain, with an open boundary in one direction and periodic boundaries in the other two, is calculated using the fast Fourier transformation (FFT) method developed by \cite{2009ApJ...693.1316K}. Ambipolar diffusion is treated in the strong coupling approximation, as described in \cite{2011ApJ...736..144B}, with super time-stepping \citep{2009ApJS..181..413C} to accelerate the evolution. 

The equations we solve are:
\begin{subequations}
\begin{align}
\frac{\partial\rho_n}{\partial t} &+ \mathbf{\nabla}\cdot\left(\rho_n\mathbf{v}\right) = 0,\\
\frac{\partial\rho_n\mathbf{v}}{\partial t} &+ \mathbf{\nabla}\cdot\left(\rho_n\mathbf{v}\mathbf{v} - \frac{\mathbf{B}\mathbf{B}}{4\pi}\right) + \mathbf{\nabla}P^* = 0,\\
\frac{\partial\mathbf{B}}{\partial t} &+ \mathbf{\nabla}\times\left(\mathbf{B}\times\mathbf{v}\right) = \mathbf{\nabla}\times\left[\frac{\left(\left(\mathbf{\nabla}\times\mathbf{B}\right)\times\mathbf{B}\right)\times\mathbf{B}}{4\pi\rho_i\rho_n\alpha}\right], \label{collision}
\end{align}
\end{subequations}
where $P^* = P + B^2/(8\pi)$. For simplicity, we adopt an isothermal equation of state $P=\rho {c_s}^2$. The numerical setup for inflow and turbulence is similar to that adopted by \hyperlink{GO11}{GO11}. For both the whole simulation box initially and the inflowing gas subsequently, we apply perturbations following a Gaussian random distribution with a Fourier power spectrum as described in \hyperlink{GO11}{GO11}. The scaling law for supersonic turbulence in GMCs obeys the relation
\begin{equation}
\frac{\delta v_\mathrm{1D} (\ell)}{\sigma_{v,\mathrm{cloud}}} = \left(\frac{\ell}{2R_\mathrm{cloud}}\right)^{1/2},
\end{equation}
where $\delta v_\mathrm{1D} (\ell)$ represents the one-dimensional velocity dispersion at scale $\ell$, and $\sigma_{v,\mathrm{cloud}}$ is the cloud-scale one-dimensional velocity dispersion. In terms of the virial parameter $\alpha_\mathrm{vir} \equiv 5{\sigma_v}^2 R_\mathrm{cloud}/ (GM_\mathrm{cloud})$ with $M_\mathrm{cloud} \equiv 4\pi \rho_0 {R_\mathrm{cloud}}^3 / 3$, and for the inflow Mach number ${\cal M}$ comparable to $\sigma_v/c_s$ of the whole cloud, the three-dimensional velocity dispersion $\delta v=\sqrt{3}\cdot\delta v_\mathrm{1D}$ at the scale of the simulation box would be
\begin{equation}
\delta v (L_\mathrm{box}) = \sqrt{3} \left(\frac{\pi G \alpha_\mathrm{vir}}{15}\right)^{1/4} {\cal M}^{1/2} {c_s}^{1/2} {\rho_0}^{1/4} {L_\mathrm{box}}^{1/2}.
\end{equation}
To emphasize the influence of the cloud magnetization instead of the perturbation field, our simulations are conducted with $10\%$ of the value $\delta v (L_\mathrm{box})$,
or $\delta v = 0.14$~km/s with $\alpha_\mathrm{vir} = 2$.
With larger $\delta v (L_\mathrm{box})$, simulations can still form cores, but because non-self-gravitating clumps can easily be destroyed by strong velocity perturbations and no core can form before the turbulent energy dissipates, it takes much longer, with corresponding higher computational expense.

\subsection{Model Parameters}
\label{sec::parameters}

\begin{figure}[t]
\begin{center}
\includegraphics[scale=0.4]{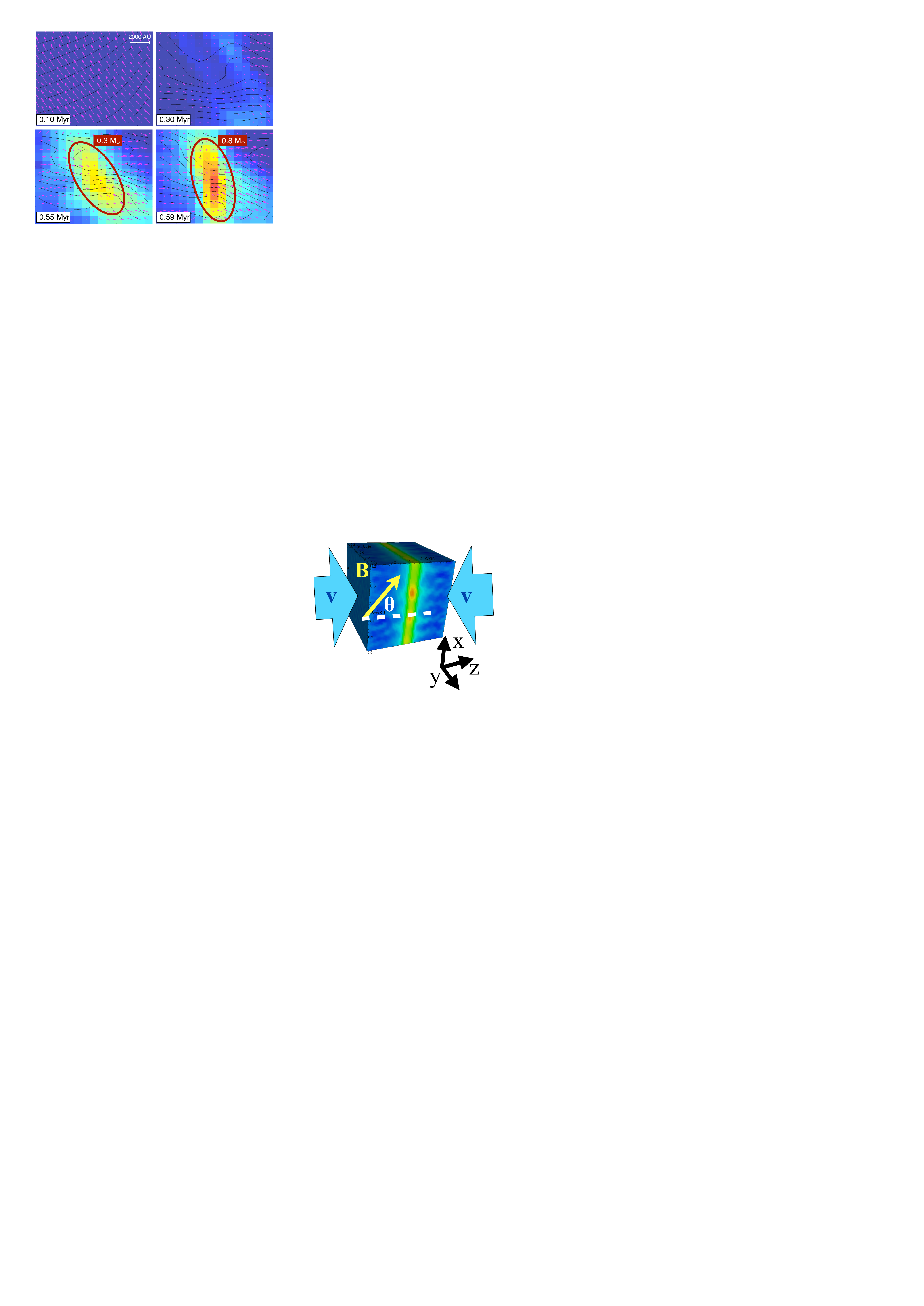}
\caption{The schematic configuration for our simulations.}
\label{setup}
\end{center}
\end{figure}

A schematic showing our model set-up is shown in Figure~\ref{setup}. Our simulation box is $1$~pc on each side and represents a region within a GMC where a large-scale supersonic converging flow with velocity $\mathbf{v}_0$ and $-\mathbf{v}_0$ (i.e. in the center-of-momentum frame) collides. The $z$-direction is the large-scale inflow direction, and we adopt periodic boundary conditions in the $x$- and $y$-directions. We initialize the background magnetic field in the cloud, $\mathbf{B}_0$, in the $x$-$z$ plane, with an angle $\theta$ with respect to the convergent flow. For simplicity, we treat the gas as isothermal at temperature $T=10$~K, such that the sound speed is $c_s=0.2$~km/s. The neutral density within the cloud, $\rho_0$, is set to be uniform in the initial conditions and in the upstream converging flow. 

It has been shown that ionization-recombination equilibrium generally provides a good approximation to the ionization fraction within GMCs for the regime under investigation (\hyperlink{CO12}{CO12}). Thus, the number density of ions in our model can be written as
\begin{equation}n_i = \frac{\rho_i}{\mu_i}= 10^{-6} \chi_{i0} \left(\frac{\rho_n}{\mu_n}\right)^{1/2},
\label{rec-ion}
\end{equation}
with
\begin{equation}
\chi_{i0} \equiv 10^6 \times \sqrt{\frac{\zeta_\mathrm{CR}}{\alpha_\mathrm{gas}}}
\label{chiDef}
\end{equation}
determined by the cosmic-ray ionization rate ($\zeta_\mathrm{CR}$) and the gas-phase recombination rate ($\alpha_\mathrm{gas}$). The ionization coefficient, $\chi_{i0}$, has values $\sim 1-20$ \citep{2010ApJ...720.1612M}, and is the model parameter that controls ambipolar diffusion effects in our simulations, following \hyperlink{CO12}{CO12}. We use typical values of the mean neutral and ion molecular weight $\mu_n$ and $\mu_i$ of $2.3 m_\mathrm{H}$ and $30 m_\mathrm{H}$, respectively, which give the collision coefficient (see Equation~(\ref{collision})) between neutrals and ions $\alpha = 3.7\times 10^{13}$~cm$^3$s$^{-1}$g$^{-1}$.

The physical parameters defining each model are $\rho_0$, $v_0 = |\mathbf{v}_0|$, $B_0 = |\mathbf{B}_0|$, $\theta$, and $\chi_{i0}$. We set the upstream neutral number density to be $ n_0 = \rho_0 / \mu_n = 1000$~cm$^{-3}$ in all simulations, consistent with typical mean molecular densities within GMCs\footnote{Note that the upstream neutral number density we adopted here is $n_0 = n_{\mathrm{neutral},0} \equiv n_\mathrm{H_2} + n_\mathrm{He} = 0.6 n_\mathrm{H} = 1.2 n_\mathrm{H_2}$, with GMC observations giving $n_\mathrm{H_2} \sim 10^2-10^3$~cm$^{-3}$. Also note that $\mu_n \equiv \rho_n/n_n = (\rho_\mathrm{H_2} + \rho_\mathrm{He}) / (n_\mathrm{H_2} + n_\mathrm{He}) = (0.5n_\mathrm{H} \times 2m_\mathrm{H} + 0.1n_\mathrm{H} \times 4m_\mathrm{H})/(0.5 n_\mathrm{H} + 0.1 n_\mathrm{H}) =  2.3 m_\mathrm{H}$.} \citep[e.g.][]{1981MNRAS.194..809L,2000prpl.conf...97W,2007A&A...471..103B,2008ApJ...686..948B}. We choose the upstream $B_0 = 10$~$\mu$G as typical of GMC values \citep{1989ApJ...338L..61G,1993ApJ...407..175C,2005LNP...664..137H,2005ApJ...624..773H} for all our simulations. To keep the total number of simulations practical, we set the large-scale inflow Mach number to ${\cal M}=10$ for all models. Exploration of the dependence on Mach number of ambipolar diffusion and of core formation has been studied in previous simulations (\hyperlink{CO12}{CO12} and \hyperlink{GO11}{GO11}, respectively). For our parameter survey, we choose $\theta=5$, $20$, and $45$~degrees to represent small ($\theta<\theta_\mathrm{crit}$), intermediate ($\theta>\theta_\mathrm{crit}$), and large ($\theta\gg\theta_\mathrm{crit}$) angles between the inflow velocity and cloud magnetic field. For each $\theta$, we conduct simulations with $\chi_{i0} = 3$, $10$, and ideal MHD to cover situations with strong, weak, and no ambipolar diffusion. We also run corresponding hydrodynamic simulations with same $\rho_0$ and $v_0$ for comparison.

A full list of models is contained in Table~\ref{modelpar}. Table~\ref{modelpar} also lists the steady-state post-shock properties, as described in Section~\ref{sec: obshock}. Solutions for all three types of shocks are listed for the $\theta=5^\circ$ (A5) case. For the $\theta=20^\circ$ and $\theta=45^\circ$ cases, there is only one shock solution. Also included in Table~\ref{modelpar} are the nominal values of critical mass and radius for spherically symmetric volumes to be self-gravitating under these steady-state post-shock condition, as discussed in Section~\ref{spherical} (see Equations~(\ref{MthSPH}), (\ref{RthSPH}), (\ref{McritSPH}) and (\ref{RcritSPH})). Both ``thermal" and ``magnetic" critical masses are listed. In most models, $M_\mathrm{mag,sph} > M_\mathrm{th,sph}$ and $M_\mathrm{mag,sph} \gg \mathrm{M}_\odot$, indicating the post-shock regions are dominated by magnetic support, and either ambipolar diffusion or anisotropic condensation would be needed to form low-mass supercritical cores, as discussed in Section~\ref{spherical}. On the other hand, the \textit{quasi-hydrodynamic} shock solution for models with $\theta<\theta_\mathrm{crit}$ (i.e.~A5 cases) has $M_\mathrm{mag,sph}<M_\mathrm{th,sph}<\mathrm{M}_\odot$ downstream. If this shock solution could be sustained, then in principle low-mass supercritical cores could form by spherical condensation of post-shock gas.

\begin{table}[t]
\small
\begin{center}
  \begin{threeparttable}
\caption{Summary of the simulation model parameters.}
\label{modelpar}
\vspace{.1in}
\begin{tabular}{ l || c c c | c c c | c c c c}
  \hline
 \multirow{3}{*}{Model} & \multicolumn{3}{c|}{model settings\tablenotemark{\wedge}} & \multicolumn{3}{c|}{steady-state post-shock solutions} & \multicolumn{4}{c}{gravitational critical scales\tablenotemark{\S}}\\
  \cline{2-11}
    & $\theta$ & $\chi_{i0}$ & $B_\perp$ & $n_\mathrm{ps}$ & $B_\perp$ & $B_\mathrm{tot}$ & $M_\mathrm{th,sph}$ & $R_\mathrm{th,sph}$ & $M_\mathrm{mag,sph}$ & $R_\mathrm{mag,sph}$ \\ 
   & (deg) & & ($\mu$G) & ($10^4$~cm$^{-3}$) & ($\mu$G) & ($\mu$G) & (M$_\odot$) & (pc) & (M$_\odot$) & (pc) \\
  \hline
  HD\tablenotemark{\P} & $-$ & $-$ & $-$ & 10.0 & $-$ & $-$ & 0.44 & 0.03 & $-$ & $-$ \\
  \hline
  A5X3 & 5 & 3 & 0.87 & & & & & \\
  A5X10 & 5 & 10 & 0.87 & & & & & \\
  \multirow{3}{*}{A5ID\tablenotemark{*}\tablenotemark{\ddagger}} & \multirow{3}{*}{5} & \multirow{3}{*}{$-$\tablenotemark{*}} & \multirow{3}{*}{0.87} & 1.51 & 55.3 & 56.2 & 1.14 & 0.07 & 11 & 0.15 \\
  & & & & 8.93\tablenotemark{\dagger} & -20.2\tablenotemark{\dagger} & 22.5\tablenotemark{\dagger} & 0.47\tablenotemark{\dagger} & 0.03\tablenotemark{\dagger} & 0.01\tablenotemark{\dagger} & 0.01\tablenotemark{\dagger} \\
  & & & & 2.79 & -51.8 & 52.7 & 0.84 & 0.05 & 2.6 & 0.07 \\
  \hline
  A20X3 & 20 & 3 & 3.42 & & & & &\\
  A20X10 & 20 & 10 & 3.42  & & & & &\\
  A20ID\tablenotemark{*} & 20 & $-$\tablenotemark{*} & 3.42 & 0.96 & 56.0 & 56.7 & 1.43 & 0.08 & 28 & 0.23 \\
  \hline
  A45X3 & 45 & 3 & 7.07 & & & & &\\
  A45X10 & 45 & 10 & 7.07 & & & & &\\
  A45ID\tablenotemark{*} & 45 & $-$\tablenotemark{*} & 7.07 & 0.69 & 57.9 & 58.3 & 1.68 & 0.10 & 59 & 0.33 \\
  \hline 
\end{tabular}
    \begin{tablenotes}
      \footnotesize
      \item $^\wedge$In the model settings, $\theta$ is the angle between inflow velocity and the magnetic field, and $B_\perp$ is the upstream magnetic field perpendicular to the shock front.
      \item $^\S$The critical masses and sizes for a spherical core at ambient post-shock conditions to have gravity exceed thermal or magnetic forces, calculated from Equation~(\ref{MthSPH}), (\ref{RthSPH}), (\ref{McritSPH}), and (\ref{RcritSPH}).
      \item $^\P$Hydrodynamics; no magnetic field, or $\chi_{i0}=0$.
      \item $^\ast$Ideal MHD; neutrals and ions are perfectly coupled.
      \item $^\ddagger$The A5 model satisfies $\theta < \theta_\mathrm{crit}$ and has three shock solutions (see Section~\ref{sec: obshock}). We list all three. The post-shock conditions in simulations may be a combination of these possible solutions.
      \item $^\dagger$The quasi-hydrodynamic solution.
    \end{tablenotes}
  \end{threeparttable}
\end{center}
\end{table}

In order to collect sufficient statistical information on the core properties from simulations, we repeat each parameter set $6$ times with different random realizations of the same perturbation power spectrum for the turbulence. The resolution is $256^3$ for all simulations such that $\Delta x \approx 0.004$~pc, or $\sim 800$~AU. We tested this setup with two times of this resolution ($\Delta x \approx 0.002$~pc), and the resulting dense structures are highly similar. Though the individual core properties vary around $\pm 50\%$, the median values (which are more important in our statistical study) only change within $\pm 10-30\%$. Thus, our simulations with $\Delta x\approx 0.004$~pc are well-resolved for investigations of core properties.

\subsection{Analysis of Core Properties}
\label{sec: MagFluxCal}

To measure the physical properties of the cores formed in our simulations, we apply the \textit{GRID} core-finding method developed by \hyperlink{GO11}{GO11}, which uses gravitational potential isosurfaces to identify cores. In this approach, the largest closed potential contour around a single local minimum of the gravitational potential defines the material eligible to be part of a core. We define the bound core region as all the material within the largest closed contour that has the sum of gravitational, magnetic, and thermal energy negative.\footnote{The gravitational, thermal, and magnetic energy density in each zone are $u_g = -\rho\Delta\Phi_g$, $u_\mathrm{th} = 3nkT/2$, and $u_B = B^2/8\pi$, respectively, where $\Delta\Phi_g$ is the difference in gravitational potential relative to the largest closed contour, and $n$ is the neutral number density defined as $n=\rho/\mu_n$. The self-gravitating core consists of all zones with $u_g + u_\mathrm{th} + u_B < 0$.} All of our cores are, by definition, self-gravitating.

The essential quantity to measure the significance of magnetic fields in self-gravitating cores is the ratio of mass to magnetic flux \citep{1956MNRAS.116..503M,1976ApJ...210..326M}. From Gauss's law the net flux of the magnetic field through a closed surface is always zero. As a result, to measure the magnetic flux within a core, we need firstly to define a cross-section of the core, and then measure the net magnetic flux through the surface of the core defined by this cross-section (which is the same as the flux through the cross-section itself). 

To define the cross-section through a core, we use the plane perpendicular to the average magnetic field that also includes the minimum of the core's gravitational potential. This choice ensures that we measure the magnetic flux through the part of the core with strongest gravity. After defining this plane, we separate the core into an upper half and a lower half, and measure the magnetic flux $\Phi_B$ through one of the halves. In practice, we compute this by firstly finding all zones that contain at least one face which is on the core surface, and assign normal vectors $\mathbf{{\hat n}}$ (pointing outwards) to those faces. From these, we select only those in the upper ``hemisphere" of the core. After we have a complete set of those grid-faces that are on the upper half of the core surface, we sum up their $\mathbf{B}\cdot \mathbf{{\hat n}}$ to get the net magnetic flux of the core. This method is tested in spherical and rectangular `cores' with magnetic fields in arbitrary directions. Note that this method works best when the core is approximately spherical (without corners).

After we have the measurement of magnetic flux $\Phi_B$, we can calculate the mass-to-flux ratio of the core, $M/\Phi_B$. This determines whether the magnetic field can support a cloud against its own self-gravity. The critical value of $M/\Phi_B$ differs with the geometry of the cloud, but the value varies only within $\sim 10\%$ \citep[e.g.][or see review in \citealt{2007ARA&A..45..565M}]{1976ApJ...210..326M,1978PASJ...30..671N,1988ApJ...335..239T}. We therefore choose the commonly used value $(2\pi\sqrt{G})^{-1}$ (e.g. \citealt{2011ApJ...728..123K}; \citealt{2011MNRAS.414.2511V}; \hyperlink{CO12}{CO12}) as a reference value, and define the normalized mass-to-flux ratio as $\Gamma \equiv  2\pi\sqrt{G}\cdot M/\Phi_B$ (see Equation~(\ref{GammaDef})). For a prestellar core with $\Gamma > 1$, the gravitational force exceeds the magnetic support and the core is magnetically supercritical. A subcritical core has $\Gamma < 1$ and is ineligible for wholesale collapse unless magnetic fields diffuse out.

\section{Sample Evolution of Structure}
\label{sec::evolution}

\begin{figure}
\begin{center}
\includegraphics[scale=0.12]{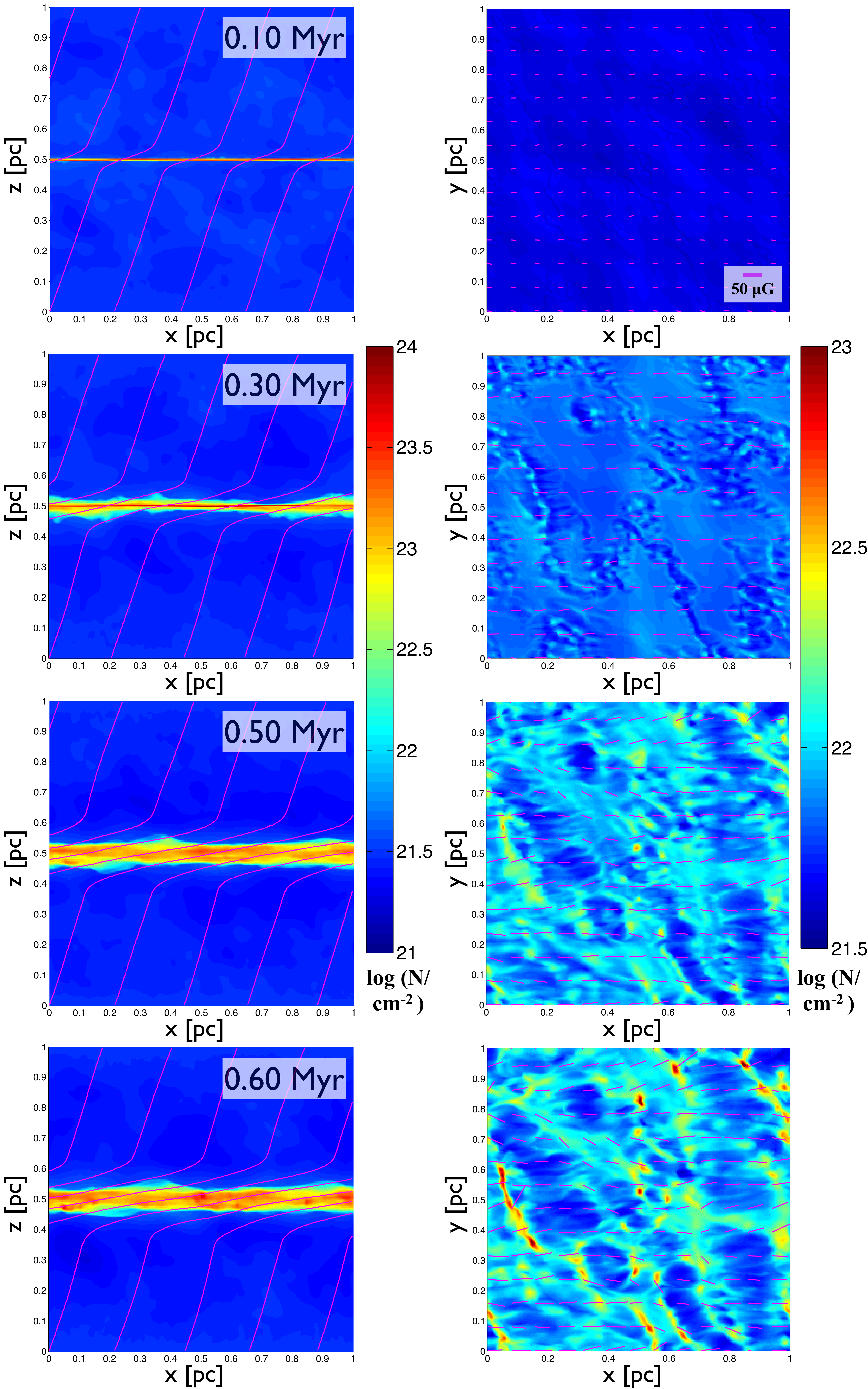}
\caption{An example of the evolution of the column density (colormap) and magnetic field structures (pink lines on left and segments on right) projected to the $x$-$z$ plane \textit{(left panel)} and $x$-$y$ plane \textit{(right panel)}, for model A20X10. Magnetic fields (integrated over the whole box) bend through the shocked gas layer, as seen on left. Right panel shows $x$-$y$ projections (with segment lengths indicating strength) of the magnetic field, which points primarily from left to right. The box size is $(1\mathrm{pc})^3$. }
\label{Evo1}
\end{center}
\end{figure}

\begin{figure}
\begin{center}
\includegraphics[scale=0.13]{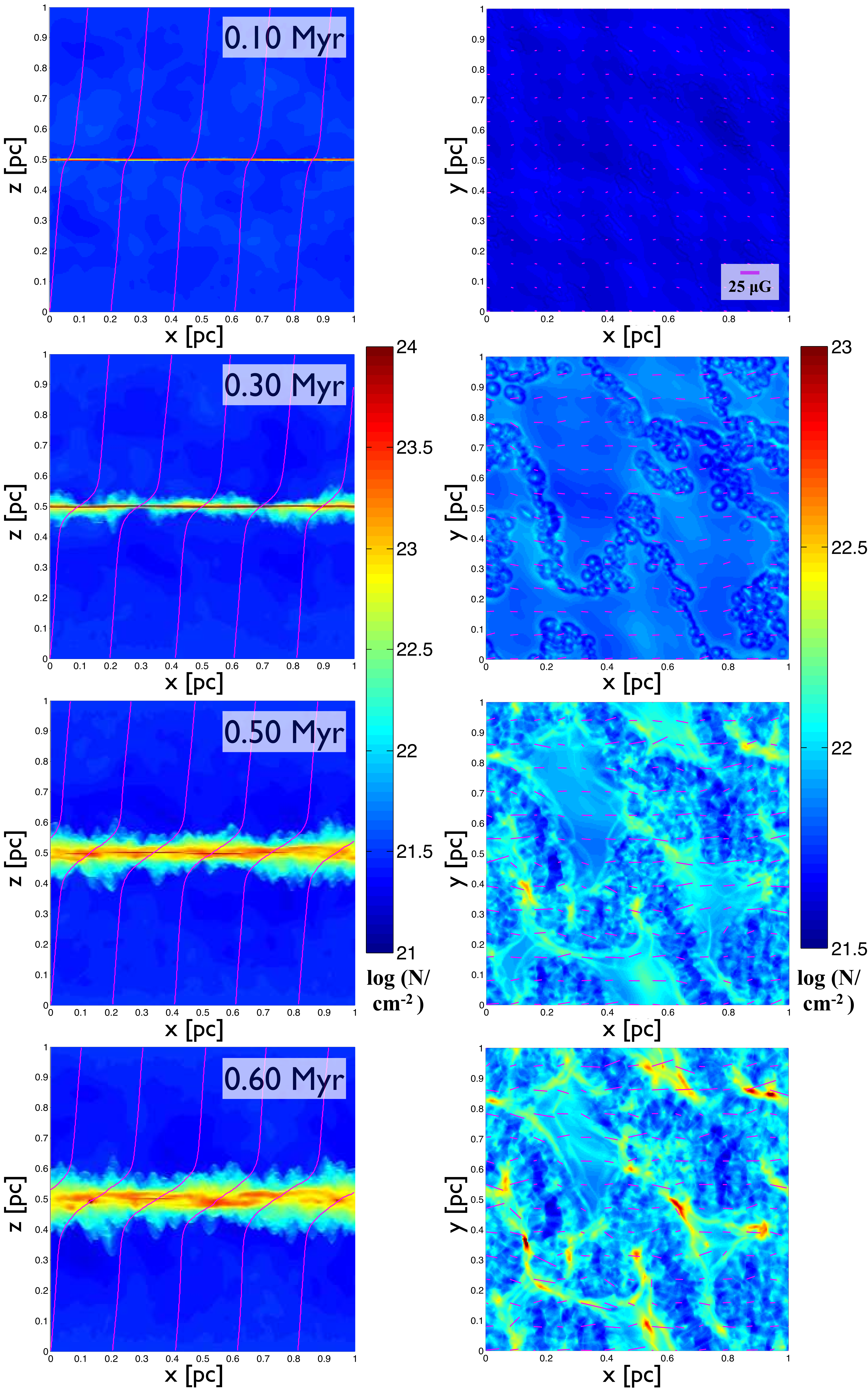}
\caption{Similar to Figure~\ref{Evo1}, but for model A5X3 with upstream magnetic field nearly parallel to the inflow ($\theta=5^\circ$), and low ionization fraction ($\chi_{i0}=3$).}
\label{Evo2}
\end{center}
\end{figure}

\begin{figure}
\begin{center}
\includegraphics[scale=0.13]{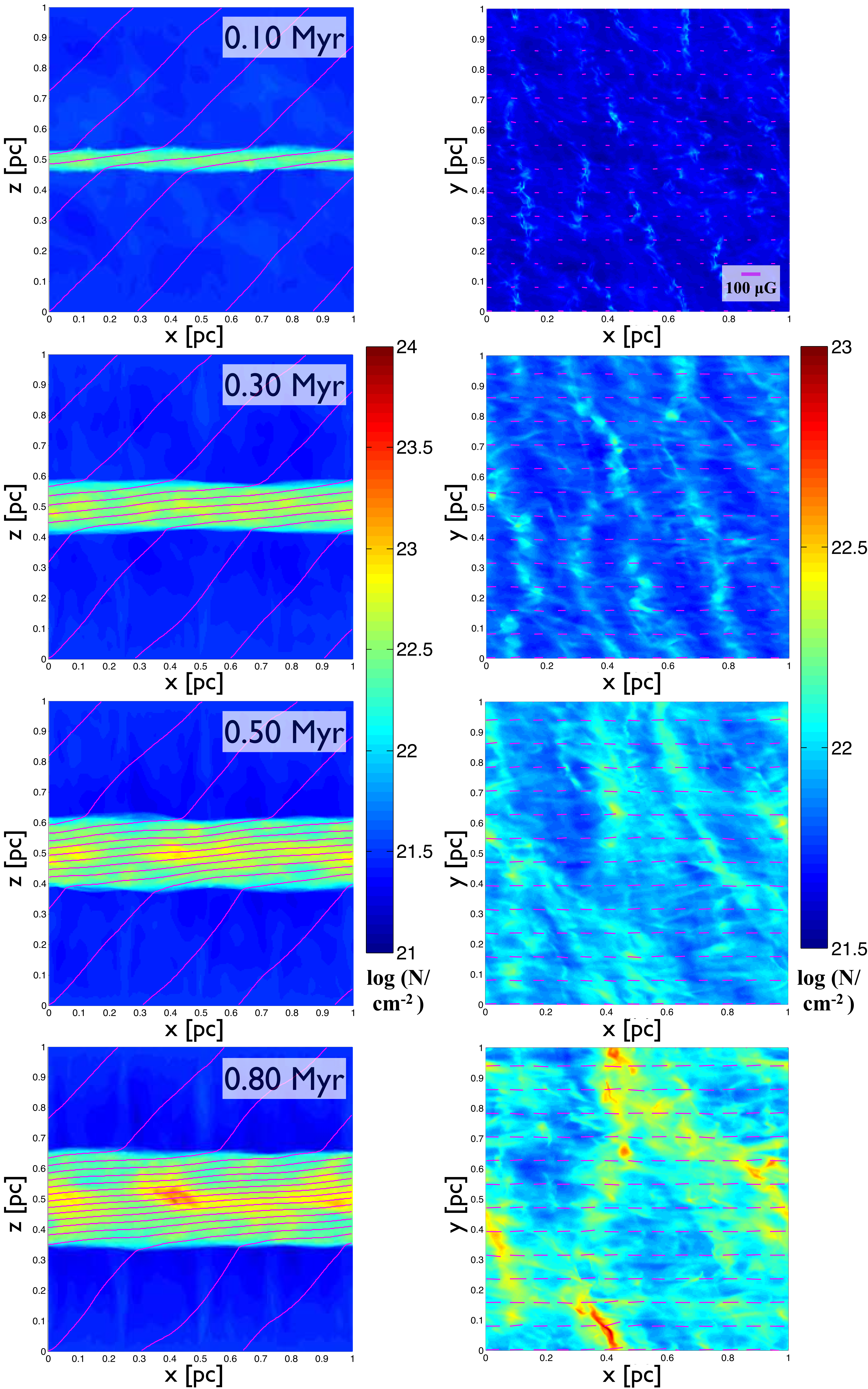}
\caption{Similar to Figure~\ref{Evo1}, but for model A45ID, with $45^\circ$ angle between upstream $\mathbf{v}$ and $\mathbf{B}$, and ideal MHD. }
\label{Evo3}
\end{center}
\end{figure}

Figure~\ref{Evo1} shows typical evolution of column density and magnetic field\footnote{The magnetic field lines shown in left panels of Figure~\ref{Evo1}, \ref{Evo2}, and \ref{Evo3} are contours of the absolute value of the magnetic vector potential $\mathbf{\Psi}$ in the direction perpendicular to the plane plotted. By definition, $\mathbf{B} = \nabla\times\mathbf{\Psi}$, and therefore $B_x = d\Psi_z/dy$, $B_y=-d\Psi_z/dx$. If we start with $\Psi_z=0$ in the lower-left corner ($x=y=0$), we can compute $\Psi_z(0,y) = \int_0^y B_x(0,y')dy'$, and $\Psi_z(x,y)=\Psi_z(0,y)-\int_0^x B_y(x',y)dx'$. After we have $\Psi_z$ everywhere, we make contours to show the magnetic field structures, with fixed spacing so $\delta\Psi=$constant.} in our numerical simulations. The simulations start with uniform density and constant magnetic field. When compressed by the supersonic converging flows, the magnetic fields perpendicular to the converging flows are amplified in the post-shock dense region. Seeded by turbulent velocity perturbations, dense structures form within the compressed layer. 

The post-shock structure can be very different for different model parameters. Figure~\ref{Evo2} and \ref{Evo3} provide examples with weak (small $\theta$ and/or small $\chi_{i0}$) and strong (large $\theta$ and/or large $\chi_{i0}$) magnetic effects in the shocked gas. The thickness of the post-shock layer is very different for these two extreme cases. Especially at early time ($0.3$~Myr), structure is also different in these two cases, with stronger magnetic effects producing filaments perpendicular to the magnetic field. The timescale at which compressed layers become gravitationally unstable and start to form cores also differ. Note that in the cases with ambipolar diffusion (Figure~\ref{Evo1} and \ref{Evo2}), a highly-compressed layer forms in the center of the post-shock region. Quantitatively, we measured the average density within the $z=0.5~\mathrm{pc} \pm \Delta x$ layer at $t=0.3~\mathrm{Myr}$ for model A5X3, and found this overdense layer has $\overline{n} \approx 1.4\times10^5$~$\mathrm{cm}^{-3}$, which exceeds the steady-MHD shock jump condition predicted in Table~\ref{modelpar} even for the {\it quasi-hydrodynamic} solution. This is a direct evidence of the existence of transient stage of ambipolar diffusion (\hyperlink{CO12}{CO12}).

Table~\ref{simresults} lists the physical properties of the post-shock layers measured at $t=0.2$~Myr as well as the corresponding values of the critical mass and size of a spherical region under these ambient conditions. Generally, models with upstream magnetic field almost parallel to the inflow (A5 models) have weaker post-shock magnetic field than that for a \textit{fast} shock (see Table~\ref{modelpar}) even with ideal MHD (A5ID), indicating that the \textit{quasi-hydrodynamic} shock mode discussed in Section~\ref{theory} plays a role.
Also, models with stronger transient ambipolar diffusion effect (smaller $\chi_{i0}$) have higher density and weaker magnetic field in the post-shock layer, and thus it would be easier to form self-gravitating cores promptly (small $M_\mathrm{th,sph}$ and $M_\mathrm{mag,sph}$ values). 

The difference in post-shock magnetic field among models with same upstream magnetic obliquity but various ionization levels can be explained by varying transient ambipolar diffusion. From Equation~(61) in \hyperlink{CO12}{CO12}, the timescale before the shock profile transitions to that of a steady C-shock is
\begin{equation}
t_\mathrm{transient} \approx \frac{2{r_f}^{1/2}}{\alpha\rho_{i,0}} = 0.34~\mathrm{Myr}\left(\frac{r_f}{10}\right)^{1/2}\left(\frac{\chi_{i0}}{10}\right)^{-1}\left(\frac{n_0}{1000~\mathrm{cm}^{-3}}\right)^{-1/2}.
\label{t_tran}
\end{equation}
Therefore, while the late-time (ideal MHD) value of $r_f$ is the same for models with same $\theta$ value, it will take $3.33$ times longer for the X3 models to reach steady-state post-shock values than the X10 models. Correspondingly, the compression rate of the magnetic field in X3 models is $0.3$ times slower than in X10 models, and thus the magnetic field within the post-shock layer is weaker in X3 models than in X10 or ideal MHD models at a given time. This tendency is clearly shown in Table~\ref{simresults}; note that since $r_f$ might be larger because of the transient ambipolar diffusion effect, the difference in post-shock magnetic field is further enhanced (smaller $\chi_{i0}$ causes higher $r_f$, resulting in longer $t_\mathrm{transient}$ and weaker $B_\mathrm{ps}$).

\renewcommand{\arraystretch}{1.1}
\begin{table}[t]
\begin{center}
  \begin{threeparttable}
\caption{Summary of the post-shock properties measured from simulations.}
\label{simresults}
\vspace{.1in}
\begin{tabular}{ l || c c c | c c c c}
  \hline
  \multirow{3}{*}{Model} & \multicolumn{3}{c|}{post-shock properties\tablenotemark{\S}} & \multicolumn{4}{c}{gravitational critical scales} \\
  \cline{2-8}
  & $\overline{n}_\mathrm{ps}$ & $\overline{B}_\mathrm{ps}$ & \multirow{2}{*}{$\overline{\beta}_\mathrm{ps}$} &$M_\mathrm{th,sph}$ & $R_\mathrm{th,sph}$ & $M_\mathrm{mag,sph}$ & $R_\mathrm{mag,sph}$ \\ 
   &  ($10^4$~cm$^{-3}$) & ($\mu$G) & & (M$_\odot$) & (pc) & (M$_\odot$) & (pc) \\
  \hline
  HD & 5.5 & $-$ & $-$ & 0.60 & 0.04 & $-$ & $-$ \\
  \hline
  A5X3 & 5.3 & 26 & 3.0 & 0.61 & 0.04 & 0.09 & 0.02 \\
  A5X10 & 5.3 & 40 & 1.3 & 0.60 & 0.04 & 0.31 & 0.03 \\
  A5ID & 2.4 & 47 & 0.43 & 0.90 & 0.05 & 2.4 & 0.08 \\
  \hline
  A20X3 & 5.3 & 45 & 1.02 & 0.61 & 0.04 & 0.45 & 0.03 \\
  A20X10 & 3.6 & 68 & 0.30 & 0.74 & 0.04 & 3.4 & 0.07 \\
  A20ID & 1.4 & 78 & 0.09 & 1.2 & 0.07 & 33 & 0.22 \\
  \hline
  A45X3 & 4.2 & 60 & 0.45 & 0.69 & 0.04 & 1.7 & 0.06 \\
  A45X10 & 2.7 & 86 & 0.14 & 0.85 & 0.05 & 12 & 0.13 \\
  A45ID & 0.91 & 96 & 0.04 & 1.5 & 0.09 & 151 & 0.41 \\
  \hline 
\end{tabular}
    \begin{tablenotes}
      \footnotesize
      \item $^\S$Post-shock properties are measured at $t=0.2$~Myr in each model, averaged over the whole post-shock layer. The timescale is chosen so the downstream properties are measured before the post-shock layer becomes strongly self-gravitating. 
      \end{tablenotes}
  \end{threeparttable}
\end{center}
\end{table}

\begin{figure}
\begin{center}
\includegraphics[scale=0.11]{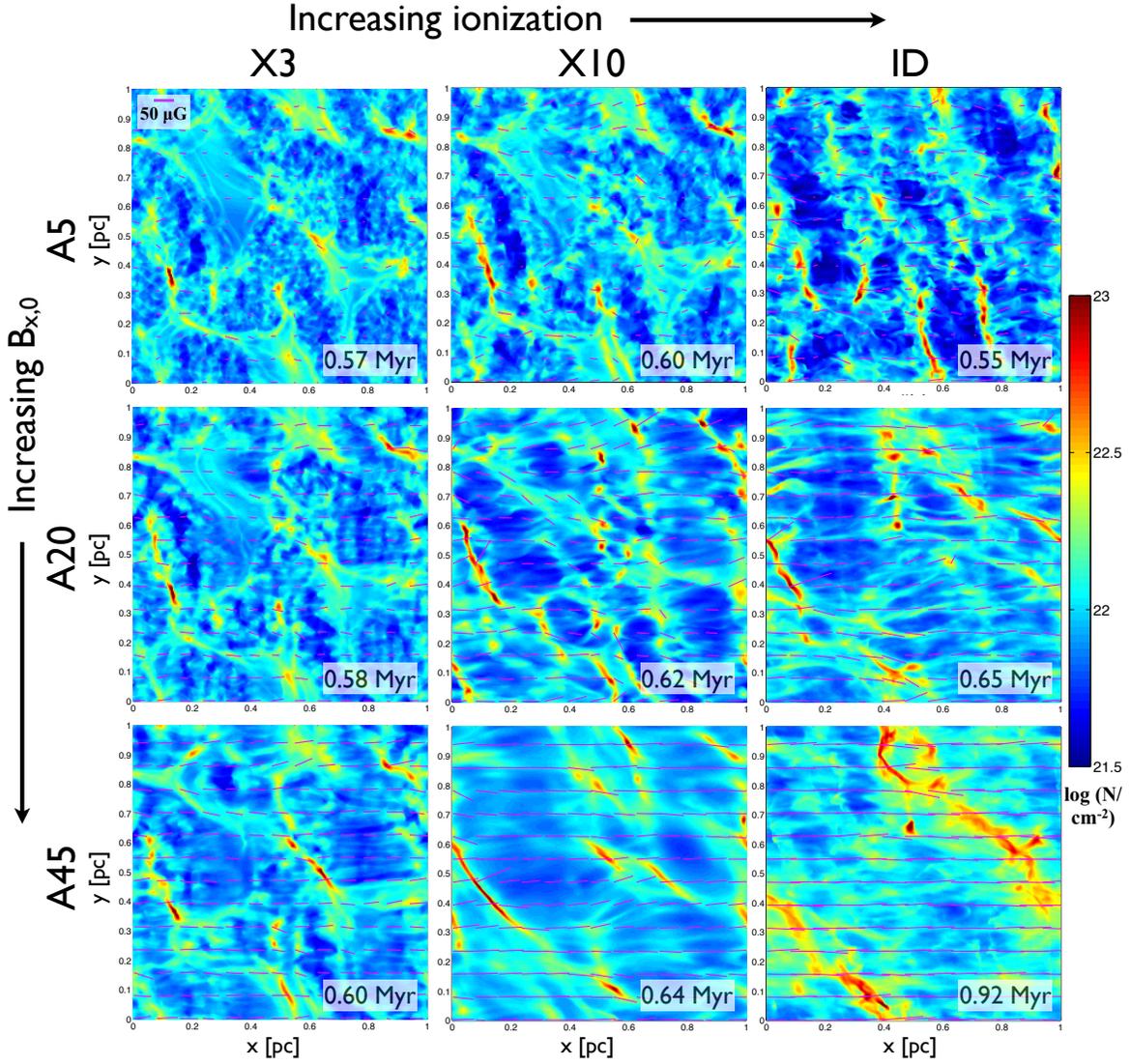}
\caption{The ``spectrum" of column density (color map) and magnetic field (pink segments) structure in the shocked gas layer for varying magnetic field parallel to the shock and ionization, at the time that maximum density reaches $10^7$~cm$^{-3}$. Model parameters are given in Table~\ref{modelpar}.}
\label{spectrum}
\end{center}
\end{figure}

Figure~\ref{spectrum} compares the density structures formed under different physical conditions, at the timescale when $n_\mathrm{max} \geq 10^7$~cm$^{-3}$ in each simulation. With low ionization (strong ambipolar diffusion), the clumps are relatively more isolated and randomly distributed, following the initial perturbation pattern. Models with high ionization (weak or no ambipolar diffusion) show well-ordered large-scale filament structures. Structures are also at larger scales for models with larger magnetic field parallel to the shock front (large $\theta$). The filaments are around $0.05$~pc wide, consistent with the observed characteristic width of filaments ($\sim 0.1$~pc, \citealp{2011A&A...529L...6A}; or see review in \citealp{2013PPVI...Andre}).
Note that the filaments are not necessary perpendicular to the magnetic field as indicated in \citet{2013ApJ...774L..31I} because the initial velocity field in our simulations is not homogeneous. 

In addition, models with moderately strong magnetization have a network of small sub-filaments aligned parallel to the magnetic field (A20X10, A20ID, A45X10, and A45ID models in Figure~\ref{spectrum}). These features are very similar to the striations identified in $^{12}$CO emission map of the Taurus molecular cloud \citep{2008ApJ...680..428G}, subsequently observed in other clouds (\citealp{2011ApJ...734...63S,2012A&A...543L...3H,2013A&A...550A..38P}; or see review in \citealp{2013PPVI...Andre}). This filament pattern is likely due to the anisotropy of turbulence at small scales in a magnetized medium \citep{1995ApJ...438..763G}, which tends to have more power for wavenumbers $\hat{k} \perp \mathbf{B}$. This leads to the formation of threads/striations/sub-filaments with small separations aligned parallel to the magnetic field in molecular clouds if the magnetic field is sufficiently strong. \citet{2003ApJ...590..858V} and \citet{2008ApJ...680..420H} found that in order to have significant turbulent anisotropy, the plasma $\beta$ must satisfy $\beta \lesssim 0.2$, which agrees with our results for when these striations are seen (see $\overline{\beta}_\mathrm{ps}$ values listed in Table~\ref{simresults}).

\section{Survey of Core Properties}
\label{sec:results}

\begin{figure}
\begin{center}
\includegraphics[scale=0.1]{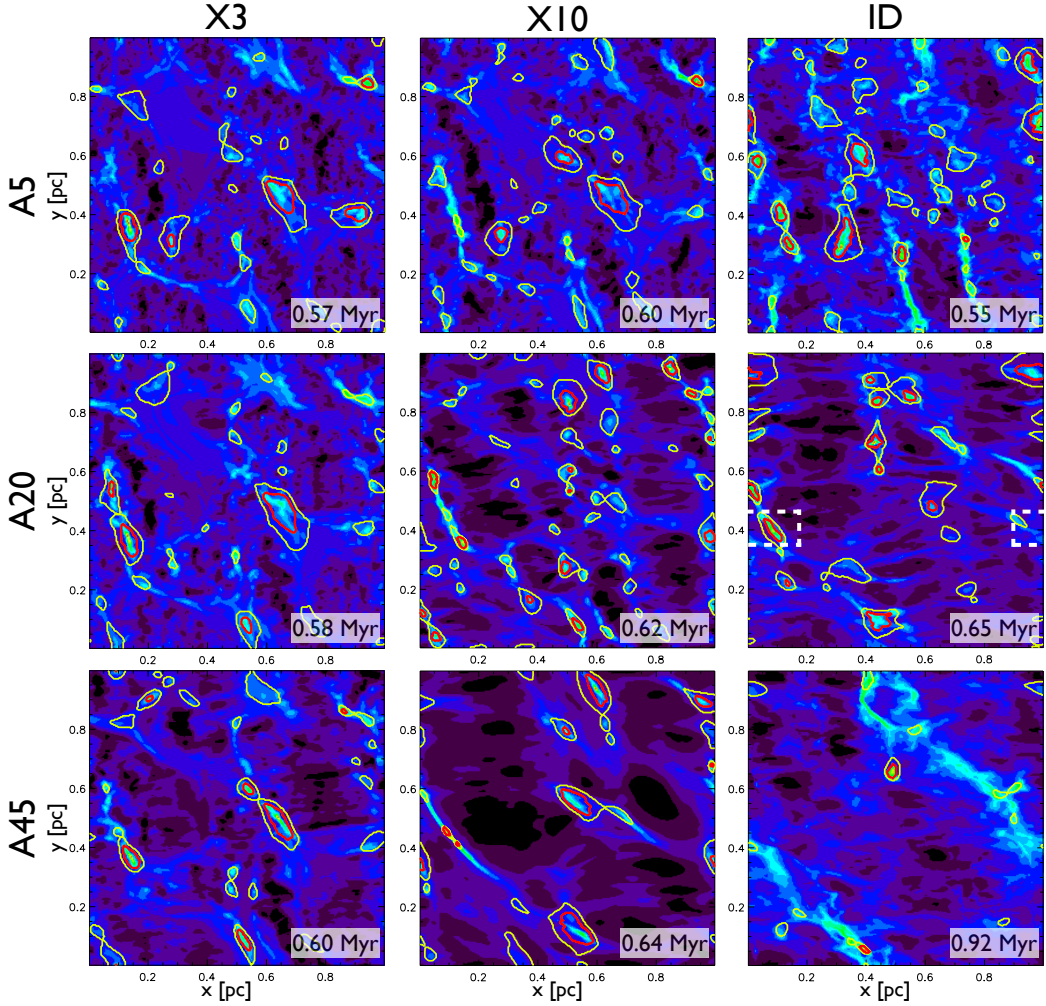}
\caption{An illustration, using one simulation for each set of model parameters, of the cores identified at the time $t_\mathrm{collapse}$ when the maximum density reaches $10^7$~cm$^{-3}$. Candidate core regions are identified using the modified \textit{GRID} core-finding method (\textit{yellow contours}), and we only consider the gravitationally bound sub-regions (\textit{red contours}). The white dashed-line box in A20ID model is the zoomed-in region shown in Figure~\ref{coreEvo} (note that the simulation box is periodic in $x$- and $y$-directions).}
\label{spectrumCore}
\end{center}
\end{figure}

We define the timescale used in Figure~\ref{spectrum} (at which $n_\mathrm{max} \geq 10^7$~cm$^{-3}$) as the moment $t_\mathrm{collapse}$ when the most evolved core starts to collapse, and measure the physical properties of all cores formed at this time. We identified hundreds of gravitationally bound cores from our $60$ simulations ($6$ runs for each parameter set), with examples illustrated in Figure~\ref{spectrumCore}. The simulation results are summarized in Table~\ref{modelsum}, including the following core properties: mean density $\overline{n}$, size $L$, mass $M$, mean magnetic field $\overline{B}$, and normalized mass-to-flux ratio $\Gamma$. To ensure the measured core properties are only for resolved structures, we omit cores with less than $27$ zones, or $L_\mathrm{core}$ smaller than $\sim 0.015$~pc. Table~\ref{modelsum} also shows for each parameter set the mean value of time $t_\mathrm{collapse}$ (at which the core properties are measured). These cores have masses, sizes, and mass-to-flux ratios similar to observed values \citep[e.g.][]{2008A&A...487..247F,2008ApJ...680..457T, 2009ApJ...699..742R,2013MNRAS.432.1424K}. 

Our results show that low-mass supercritical cores form at $t < 1$~Myr in all models: with converging velocity either nearly aligned with the magnetic field (small $\theta$) or highly oblique (large $\theta$), and for all levels of ambipolar diffusion. We also calculated the core formation efficiency (CFE) from our simulations:
\begin{equation}
\mathrm{CFE} \equiv \frac{\mathrm{mass\ in\ cores}}{\mathrm{mass\ of\ the\ shocked\ layer}} = \frac{\sum\limits_i M_{\mathrm{core}, i}}{2\rho_0v_0 t_\mathrm{collapse}\cdot L_x L_y} \approx 3.1\%.
\end{equation}
This is similar to the observed star formation efficiency (SFE), which is around $1-10\%$ \citep[e.g][]{1986ApJ...301..398M,2009ApJS..181..321E,2010ApJ...724..687L}. Note that, though the core formation timescale is slightly different from model to model (see Figure~\ref{spectrum}), the CFE does not vary significantly between models; the variance in CFE among all models is only $\sim 10\%$.

\renewcommand{\arraystretch}{1.1}
\begin{table}[t]
\begin{center}
  \begin{threeparttable}
\caption{Results from identified cores measured at $t=t_\mathrm{collapse}$.}
\label{modelsum}
\vspace{.1in}
\begin{tabular}{ l || c c c | c c c c c}
  \hline
  Model &  \# Cores & CFE\tablenotemark{\P}& $t_\mathrm{collapse}$\tablenotemark{\S} & $\overline{n}_\mathrm{core}$ & $L_\mathrm{core}$\tablenotemark{\ddagger} & $M_\mathrm{core}$ & $\overline{B}_\mathrm{core}$ & $\Gamma_\mathrm{core}$ \\ 
   & Identified\tablenotemark{\star} & (\%) & (Myr) & ($10^5$~cm$^{-3}$) & (pc) & (M$_\odot$) & ($\mu$G) & (normalized) \\
  \hline
  HD & 32 & 3.1 & 0.56 & 5.8 & 0.036 & 0.75 & $-$ & $-$ \\
  A5X3 & 40 & 3.1 & 0.58 & 5.5 & 0.032 & 0.63 & 42 & 4.4 \\
  A5X10 & 49 & 3.7 & 0.61 & 6.6 & 0.031 & 0.65 &  64 & 3.7 \\
  A5ID & 51 & 3.9 & 0.54 & 5.6 & 0.030 & 0.58 & 67 & 2.6 \\
  A20X3 & 34 & 3.0 & 0.59 & 5.6 & 0.032 & 0.72 & 60 & 3.9 \\
  A20X10 & 54 & 3.1 & 0.60 & 9.7 & 0.025 & 0.47 & 79 & 3.3 \\
  A20ID & 36 & 3.3 & 0.62 & 9.5 & 0.031 & 0.78 & 90 & 2.7 \\
  A45X3 & 42 & 3.7 & 0.60 & 8.9 & 0.031 & 0.73 & 83 & 3.7 \\
  A45X10 & 38 & 3.2 & 0.60 & 9.2 & 0.030 & 0.70 & 82 & 3.0 \\
  A45ID & 21 & 1.9 & 0.90 & 11 & 0.035 & 1.12 & 137 & 2.1 \\ 
  \hline 
\end{tabular}
    \begin{tablenotes}
      \footnotesize
      \item $^\star$We only consider gravitationally bound cores with $E_\mathrm{grav} + E_\mathrm{thermal} + E_\mathrm{B} < 0$.
      \item $^\P$CFE is the ratio of the total mass in cores to the total mass in the shocked layer at $t_\mathrm{collapse}$.
      \item $^\dagger$Columns (5)$-$(9) are averaged over all cores for each parameter set (6 simulation runs).
      \item $^\S$Collapse is defined as the time when $n_\mathrm{max} =10^7$~cm$^{-3}$ in each simulation. The $t_\mathrm{collapse}$ shown here is the mean value over all 6 runs for each parameter set.
      \item $^\ddagger$$L_\mathrm{core}$ is calculated from the total number of zones $N$ within a core, for an equivalent spherical volume: $L_\mathrm{core} = 2\times(3N/(4\pi))^{1/3}\Delta x$, where $\Delta x=1/256$~pc is the grid size.
    \end{tablenotes}
  \end{threeparttable}
\end{center}
\end{table}

\subsection{Mass and Size}
\label{MassSize}

\begin{figure}[t]
\begin{center}
\includegraphics[scale=0.7]{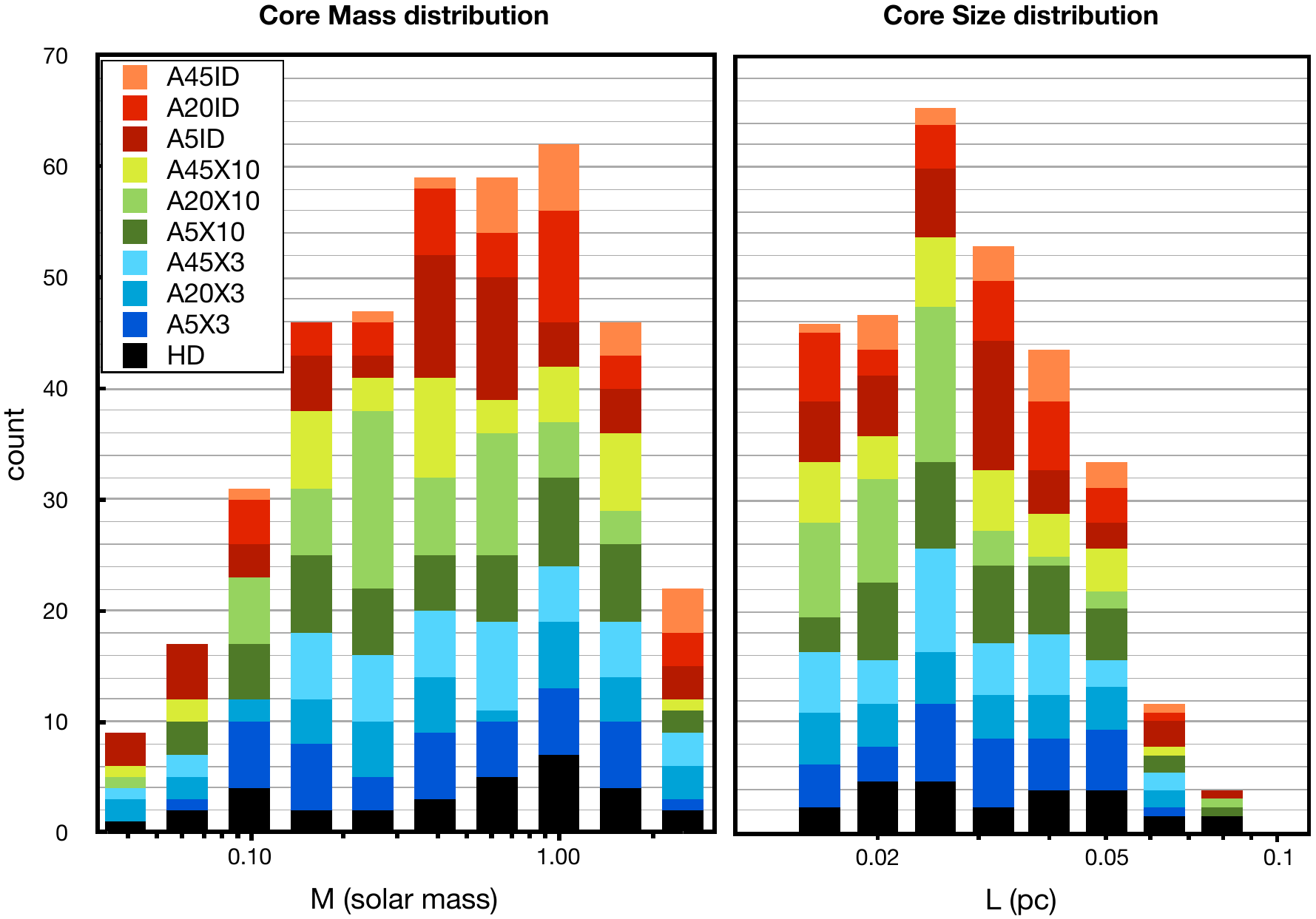}
\caption{The statistical distribution of core mass \textit{(left)} and size \textit{(right)} from all models combined.}
\label{Dist}
\end{center}
\end{figure}

\begin{figure}
\begin{center}
\includegraphics[scale=0.6]{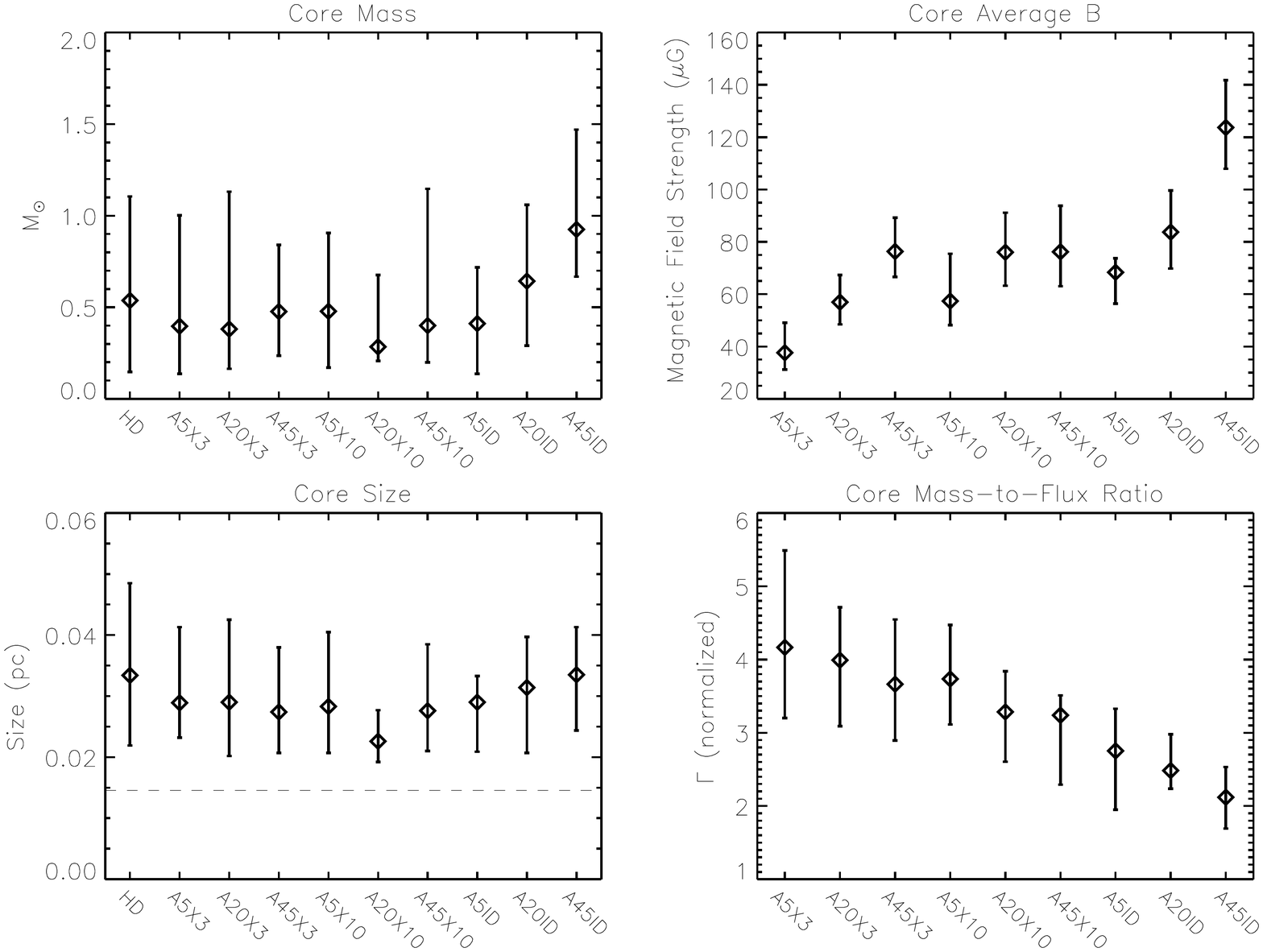}
\caption{Median (\textit{diamond}) and $\pm 25\%$ values (\textit{vertical bars}) of core mass, size, average magnetic field, and normalized mass-to-flux ratio for different model parameters. In each figure from left to right, higher X corresponds to increasing ionization, and larger A corresponds to larger angle $\theta$, or increasing pre-shock (upstream) magnetic field $B_x = B_0\sin\theta$ parallel to the shock front. The dashed line in the core size plot \textit{(bottom left)} indicates the lower limit (0.015~pc) of resolved core size; for our simulations, $0.015$~pc~$\approx3\Delta x$. }
\label{MaModel}
\end{center}
\end{figure}

Figure~\ref{Dist} shows the distribution of mass and size of cores for all model parameters. The masses range between $0.04$ to $2.5$~$M_\odot$, with peak around $\sim 0.6$~$M_\odot$; the core sizes are between $0.015 - 0.07$~pc, with peak around $\sim 0.03$~pc. These are consistent with observational results \citep[e.g.][]{2001A&A...372L..41M, 2009ApJ...691.1560I, 2009ApJ...699..742R,2013MNRAS.432.1424K}. Also, the distribution of the core mass shows a similar shape to the observed core mass function (CMF) \citep[e.g.][]{2008MNRAS.391..205S,2009ApJ...699..742R,2010A&A...518L.106K}. Interestingly, the peak in the distribution is close to value given by Equation~(7) from \hyperlink{GO11}{GO11}:
\begin{equation}
M_\mathrm{BE,\ ps} = 1.2 \frac{{c_s}^4}{\sqrt{G^3 P_\mathrm{ps}}}= 1.2 \frac{{c_s}^3}{\sqrt{G^3 \rho_0}}\frac{1}{{\cal M}} \rightarrow 0.45\ \mathrm{M}_\odot.
\end{equation}
This mass is characteristic of what is expected for collapse of a thermally-supported core that is confined by an ambient medium with pressure equal to the post-shock value\footnote{The post-shock total pressure (whether for an unmagnetized medium, as considered by \hyperlink{GO11}{GO11}, or for a magnetized medium as considered here) will be comparable to the momentum flux of the converging flow, $P_\mathrm{ps}\approx \rho_0 {v_0}^2 = \rho_0 c_s^2 {\cal M}^2$. }, where the numerical figure uses values for the mean cloud density and large-scale Mach number equal to those of the converging flow in our simulations, $n_0 = 1000$~cm$^{-3}$ and ${\cal M} = 10$. Correspondingly, since the critical ratio of mass and radius is $M_\mathrm{BE}/R_\mathrm{BE} = 2.4 {c_s}^2/G$ \citep{1956MNRAS.116..351B}, the characteristic size expected for a collapsing core formed in a post-shock region when the Mach number of the large-scale converging flow is ${\cal M}$ and the mean cloud density is $\rho_0$, is 
\begin{equation}
L_\mathrm{BE} = 2 R_\mathrm{BE} = \frac{c_s}{\sqrt{G\rho_0}}\frac{1}{{\cal M}} \rightarrow 0.04\ \mathrm{pc}.
\end{equation}
This is again comparable to the peak value of the core size distribution in Figure~\ref{Dist}.

We also separately explore the dependence of core mass, size, magnetic field strength, and mass-to-flux ratio on model parameters, as shown in Figure~\ref{MaModel}. Our results show that the core mass is relatively insensitive to both the ionization (i.e. ambipolar diffusion effect) and obliquity of the upstream magnetic field (Figure~\ref{MaModel}, top left). The median masses are within a factor 2.4 of the mean of the whole distribution, 0.68~M$_\odot$, or a factor 2 of the median of all core masses (0.47~M$_\odot$). Similarly, median core sizes vary only between 0.022~pc and 0.034~pc for the various parameter sets, with a median of 0.03~pc. Note that we chose to compare median values between different parameter sets in Figure~\ref{MaModel} instead of mean values used in Table~\ref{modelsum}, because an average can be affected by any single value being high or low relative to the other samples. The median value, on the other hand, represents the central tendency better, and with the $\pm 25\%$ values we can have a better understanding of the sample distribution. 

We note in particular that for the $\theta=20^\circ$ and $\theta=45^\circ$ ideal MHD cases, the masses in Figure~\ref{Dist} and \ref{MaModel} are more than an order of magnitude lower than the limits for a spherical region at post-shock conditions to be magnetically supercritical, as listed in Table~\ref{modelpar} and \ref{simresults}. This implies that the low-mass bound cores found in the simulations did not form isotropically. We discuss this further in Section~\ref{sec::CF}.

\begin{figure}[t]
\begin{center}
\includegraphics[scale=0.6]{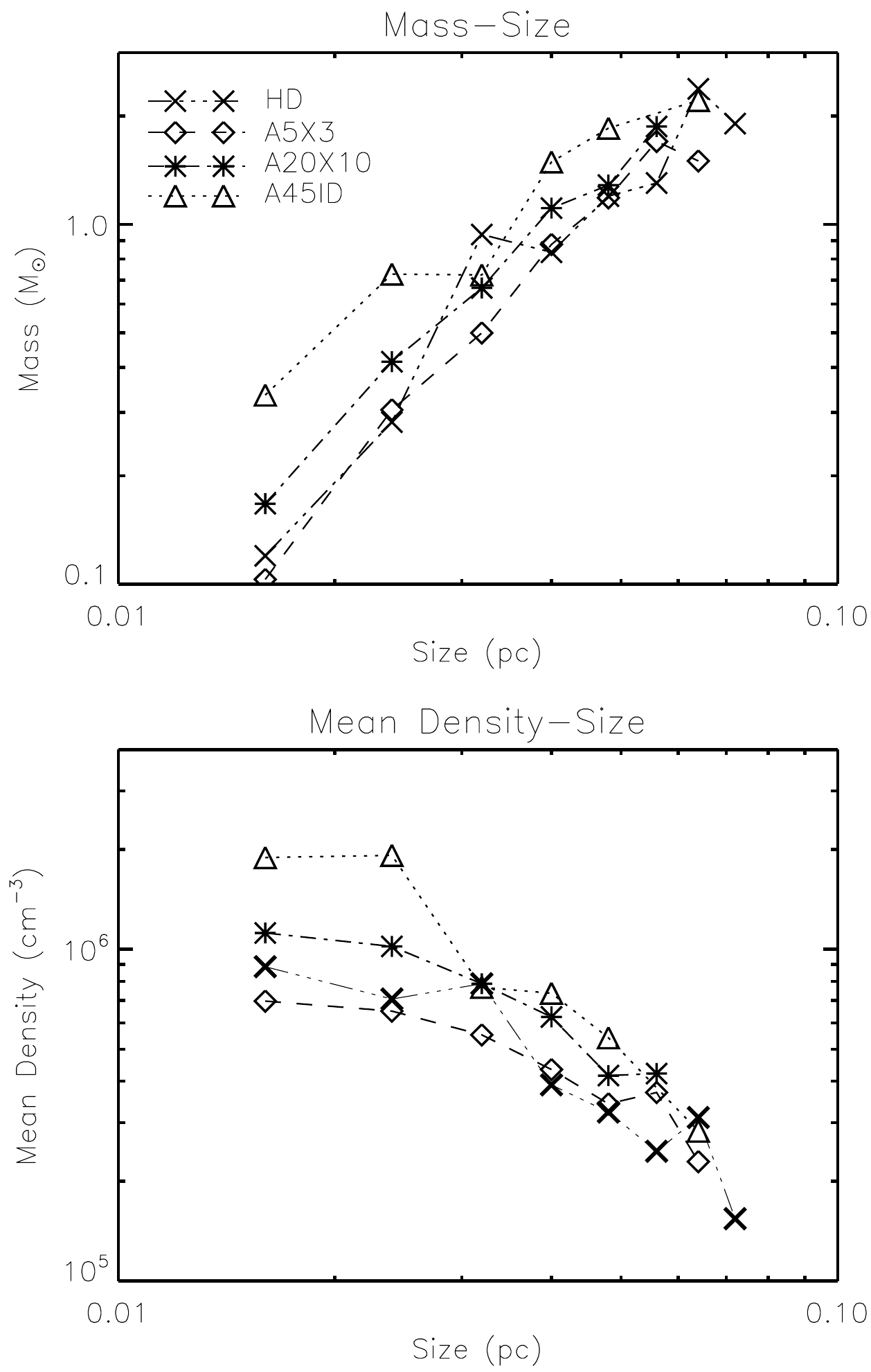}
\caption{The mass-size plot (\textit{top}) and density-size plot (\textit{bottom}) for cores identified in selected models with different magnetization and ionization levels: HD (\textit{cross}), A5X3 (\textit{diamonds}), A20X10 (\textit{asterisks}), and A45ID (\textit{triangles}).}
\label{compareML}
\end{center}
\end{figure}

To further investigate the relationship between core masses and sizes, we binned the data set by $\log L_\mathrm{core}$ and calculate the average core mass and mean density for different model parameters. The results are shown in Figure~\ref{compareML}, where we chose four models with different magnetization and ionization levels to compare: HD (hydrodynamics; no magnetization), A5X3 (low ionization, weak upstream magnetic field parallel to the shock), A20X10 (moderate ionization and magnetic field), and A45ID (ideal MHD, strong magnetic field). In both the mass-size and density-size plots, the differences among models are small, and all four curves have similar shape. In fact, from all resolved cores identified in our simulations, we found a power-law relationship between the core mass and size, $M\propto L^k$, with best-fitted value $k=2.28$. This is consistent with many core-property surveys towards different molecular clouds \citep[e.g.][]{1996ApJ...471..816E,2010MNRAS.402..603C,2010ApJ...723..492R,2013MNRAS.432.1424K}, in which $k = 1.2-2.4$ with various molecule tracers \citep[for more details, see Figure 7 and corresponding discussions in][]{2013MNRAS.432.1424K}. 

\begin{figure}[t]
\begin{center}
\includegraphics[scale=0.6]{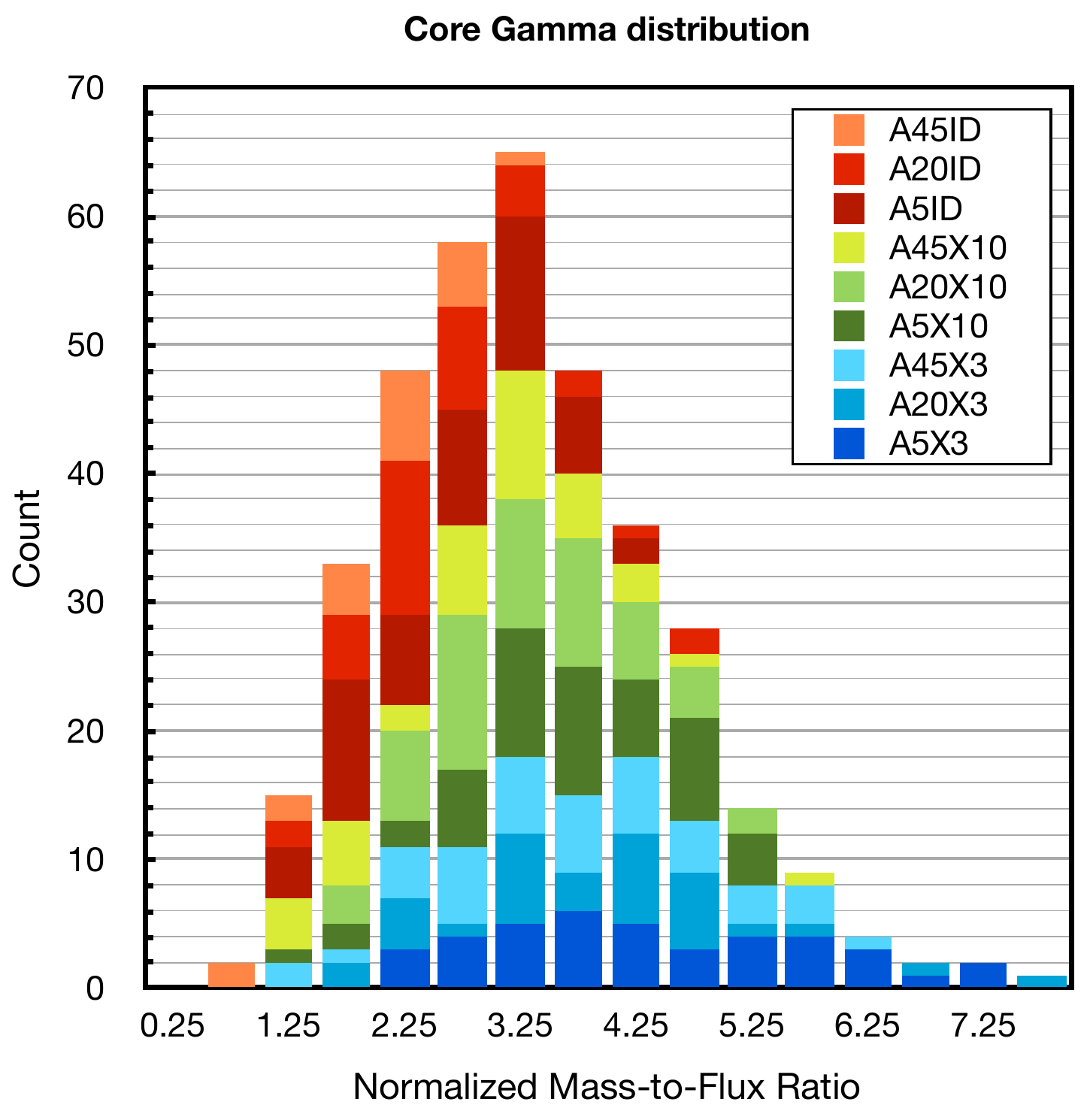}
\caption{The statistical distribution of normalized core mass-to-flux ratio $\Gamma$ from all simulations combined. Models with low ionization \textit{(blue)} preferentially have higher $\Gamma$, whereas models with ideal MHD \textit{(red)} have lower $\Gamma$.}
\label{GaDist}
\end{center}
\end{figure}

\subsection{Magnetization}
\label{magnetization}

Figure~\ref{GaDist} shows the distribution of core mass-to-flux ratio, a roughly normal distribution with peak at $\Gamma\sim 3$. This range of $\Gamma$ is quite similar to observational results \cite[$\Gamma\sim 1-4$;][]{2008A&A...487..247F,2008ApJ...680..457T}. In addition, the color-coded histogram in Figure~\ref{GaDist} shows how the mass-to-flux ratio depends on magnetization: the high-end region ($\Gamma \gtrsim 5$) is comprised of blue-green pieces (which represent models with lower ionization), while the low-end tail is mostly red and orange (highly ionized models). Note that essentially all of the cores in our simulations are magnetically supercritical ($\Gamma > 1$), which is self-consistent with our core-finding criterion (gravitationally bound; $E_g+E_\mathrm{th}+E_B<0$). 

\begin{figure}[t]
\begin{center}
\includegraphics[scale=0.8]{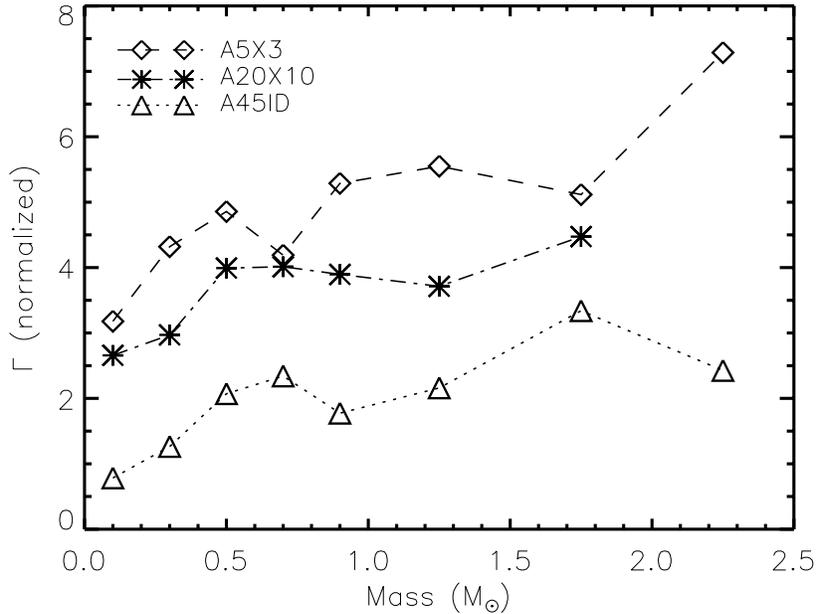}
\caption{Core mass-to-flux ratio versus core mass for sample sets of parameters.   The value of $\Gamma$ tends to increase with ionization, and to a lesser extent also increases with mass.}
\label{GammaMass}
\end{center}
\end{figure}

The tendency of models with lower ionization to form cores with higher mass-to-flux ratio is very clearly seen in Figure~\ref{MaModel} (\textit{bottom right}). The median value of the core mass-to-flux ratio also decreases with increasing $\theta$ as the value of the upstream $B_x = B_0\sin\theta$ increases.
Also from Figure~\ref{MaModel} (\textit{top right}), the average core magnetic fields show a similar tendency as in post-shock magnetic field (see Table~\ref{simresults}), which decrease at lower ionization fractions for models with same pre-shock magnetic field structure (same $\theta$).
The larger and more systematic variation of $\overline{B}$ than $M$ with model parameters suggests that the core mass-to-flux ratio is not decided by the core mass, but by the core magnetic field. This is also shown in Figure~\ref{GammaMass}, where we binned the data by $M_\mathrm{core}$ and calculated the average core mass-to-flux ratio in each bin for different models. For cores with similar mass, the mass-to-flux ratios of cores formed in environments with low ionization and magnetization are much higher than those with stronger and better coupled magnetic fields. 

The fact that the median value of magnetic field strength within the core depends on pre-shock magnetic obliquity and ionization is consistent with our discussions in Section~\ref{sec::evolution} that magnetic fields are lower in shocked regions that have longer transient timescales.
Since lower ionization fraction leads to stronger ambipolar diffusion and a longer transient stage\footnote{From \hyperlink{CO12}{CO12} and Equation~(\ref{t_tran}), the predicted duration of the transient stage is $0.3 - 1.4$~Myr for $\chi_{i0}=3$ to $10$ and our range of model parameters, assuming $r_f = r_{f,\ \mathrm{ideal\ MHD}}$.}, it is logical to expect the cores formed in weakly-ionized clouds have lower magnetic field than those formed with higher ionization fraction (or strongly-coupled ions and neutrals). 

In addition, Figure~\ref{MaModel} (\textit{top right}) shows that cores formed in models with small $\theta$ (A5 cases) have weaker magnetic fields inside even with higher ionization fraction or ideal MHD, which indicates that the magnetic field is less compressed by the shock when the inflow is almost parallel to the upstream magnetic field. This is consistent with the discussion in Section~\ref{theory}: when $\theta<\theta_\mathrm{crit}$, the MHD shock becomes a composite compounded of the regular (\textit{fast}) mode and the \textit{quasi-hydrodynamic} mode, which has relatively small magnetic field compression ratio. Thus, the magnetic obliquity relative to the shock has a similar effect to the cloud ionization fraction in determining field strengths in prestellar cores.

Based on the results shown in Section~\ref{MassSize} and \ref{magnetization}, we conclude that magnetic effects do not appear to control core mass and size. This suggests that once a core becomes strongly gravitationally bound, magnetic effects are relatively unimportant to its internal structure. However, the formation process of gravitationally bound cores is highly dependent on magnetic effects. As noted above, Figure~\ref{spectrum} shows clear differences in the large-scale structures from which cores condense; we discuss core formation further in Section~\ref{sec::CF}. Also, cores are born with either lower or higher magnetic field, depending on the magnetic field structure and the ambipolar diffusion in their surrounding environment.

\section{Anisotropic Core Formation}
\label{sec::CF}

\subsection{Examples of Simulation Evolution}

\begin{figure}
\begin{center}
\includegraphics[scale=0.25]{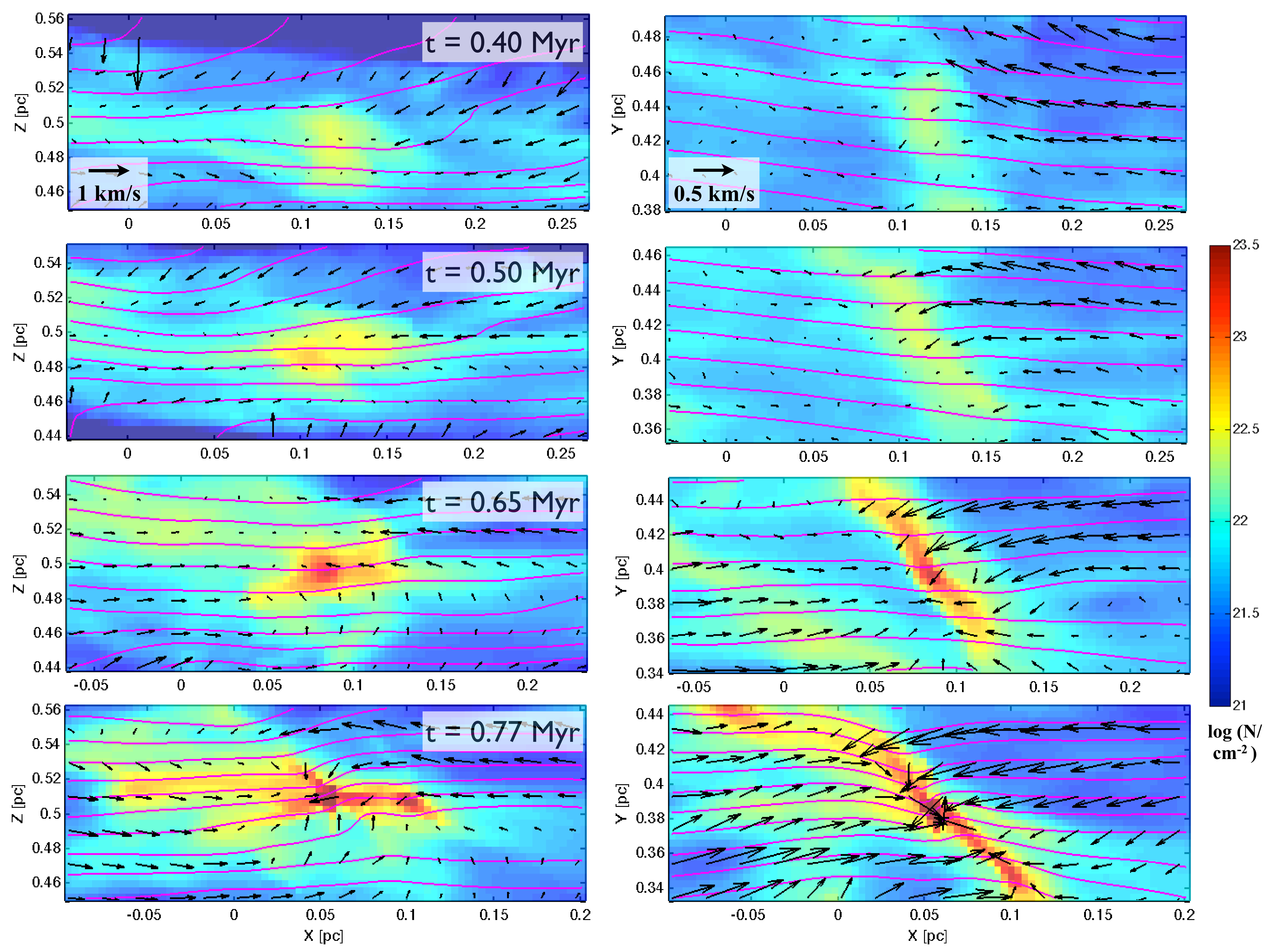}
\caption{A close-up view of magnetic field \textit{(pink lines)} and gas velocity \textit{(black arrows)} over column density \textit{(color map)} projected to the $x$-$z$ plane (\textit{left panel}) and $x$-$y$ plane (\textit{right panel}) around a forming core at different times, from model A20ID. The region shown here is indicated by the white dashed box in Figure~\ref{spectrumCore}. The size of the box $N_x\times N_y\times N_z$ is $L_\mathrm{mag,cyl} \times 4R_\mathrm{th,sph}\times 4R_\mathrm{th,sph}$, where $L_\mathrm{mag,cyl}$ and $R_\mathrm{th,sph}$ are calculated using Equation~(\ref{theoLcrit}) and (\ref{Rfinal}), respectively. The velocity vectors are density-weighted averages over the box; i.e.~$v_\mathrm{2D} (i,j) = \sum\limits_{k} (v_\mathrm{3D} (i,j,k) \rho (i,j,k)) / \sum\limits_{k} \rho (i,j,k)$. We used the same method as in the left panel in Figure~\ref{Evo1} to draw the magnetic field lines. The magnetic field line spacing and the length of the velocity vectors both indicate strength. Note that the vector scale in the right panel is $2$ times larger than in the left panel in order to better show the gas movement. The pre-shock supersonic inflows along the $z$-direction in the earlier stages (first two plots in left panel) are omitted here to focus on the post-shock dynamics. }
\label{coreEvo}
\end{center}
\end{figure}

The fact that gravitationally supercritical low-mass cores (with $M \ll M_\mathrm{mag,sph}$) can form in the highly magnetized post-shock medium even without ambipolar diffusion suggests that these cores did not contract isotropically. Figure~\ref{coreEvo} provides a close-up view of the core forming process in highly magnetized environment with ideal MHD, from model A20ID. At stages earlier than shown, the directions of the perturbed magnetic field and gas velocity are determined randomly by the local turbulence. The magnetic field is compressed by the shock (similar to Figures~\ref{Evo1}$-$\ref{Evo3}), such that in the post-shock layer it is nearly parallel to the shock front (along $\hat{\mathbf{x}}$). When the magnetic field strength increases, the velocity is forced to become increasingly aligned parallel to the flow, as shown in Figure~\ref{coreEvo}. By the time $t = 0.65$~Myr, a very dense core has formed by gathering material preferentially along the magnetic field lines. After the core becomes sufficiently massive, its self-gravity will distort the magnetic field and drag material inward even in the direction perpendicular to the magnetic field lines ($t=0.77$~Myr, Figure~\ref{coreEvo}). 
This collapsing process with a preferential direction is similar to the post-shock focusing flows found in previous studies \cite[e.g.][]{2013ApJ...774L..31I,2013MNRAS.433.1258V} where the gas is confined by the strong magnetic field in the shock-compressed region.

\begin{figure}
\begin{center}
\includegraphics[scale=0.12]{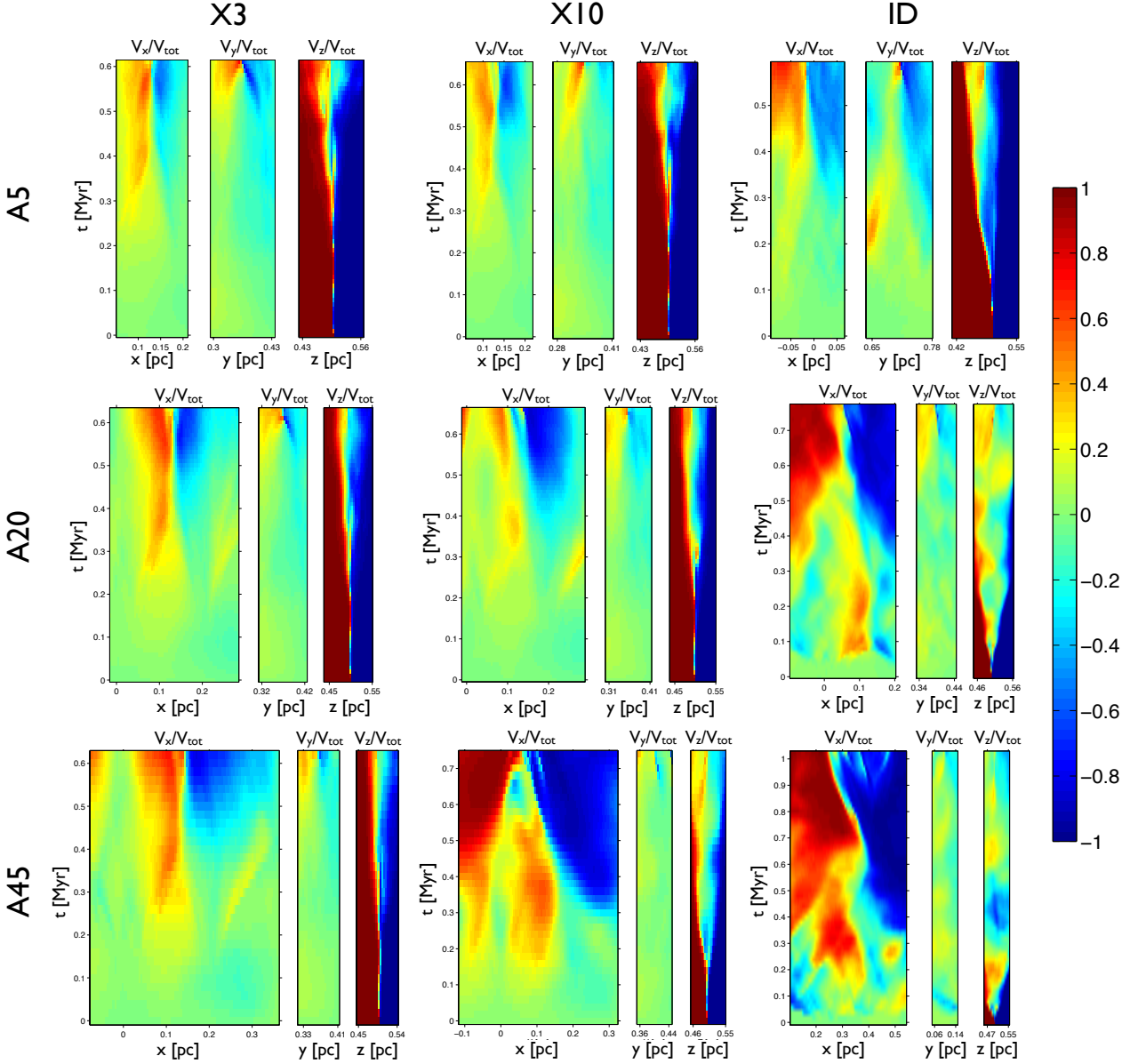}
\caption{The space-time diagrams for varying magnetic obliquity and ionization, showing the $x$- (\textit{left panels}), $y$- (\textit{middle panels}), and $z$- (\textit{right panels}) compressive components of the gas velocity averaged around a collapsing core in each model, normalized by the total velocity $v_\mathrm{tot}=\sqrt{{v_x}^2+{v_y}^2+{v_z}^2}$. The definition of box size is the same as in Figure~\ref{coreEvo}. It is evident that anisotropic condensation along the magnetic field ($x$-direction) initiates core formation in all cases.}
\label{Vtime}
\end{center}
\end{figure}

In fact, anisotropic condensation is key to core formation not only with ideal MHD, but for all cases. Figure~\ref{Vtime} shows space-time diagrams of all three velocity components ($v_x$, $v_y$, $v_z$) around collapsing cores in different parameter sets. Though models with stronger post-shock magnetic fields (larger $\theta$, larger $\chi_{i0}$, or ideal MHD) have more dominant $v_x$ (bluer/redder in the colormap), there is a general preference to condense preferentially along the magnetic field lines (in the $x$-direction) among all models, regardless of upstream magnetic obliquity and the ambipolar diffusion level. Figure~\ref{Vtime} also shows that there are flows perpendicular to the mean magnetic field (along $\mathbf{\hat{y}}$ or $\mathbf{\hat{z}}$) in the final $\sim 0.1$~Myr of the simulations, indicating the stage of core collapse. The prominent gas movement along $\mathbf{\hat{x}}$ long before each core starts to collapse shows that cores acquire masses anisotropically along the magnetic field lines, and thus anisotropic condensation is important for all models.

\subsection{Theoretical Scalings}
\label{anisotropic}

We have shown in Section~\ref{spherical} quantitatively that isotropic formation of low-mass supercritical cores is not possible for oblique shocks with ideal MHD, because the minimum mass for a spherical volume to become magnetically supercritical is $\ge 10$~M$_\odot$ (see Equation~(\ref{McritSPH}) and the $M_\mathrm{mag,sph}$ entries in Table~\ref{modelpar} and \ref{simresults} for cases A20ID and A45ID), much larger than the typical core mass ($\sim 1$~M$_\odot$). However, non-spherical regions may have smaller critical mass. Consider, for example, a core that originates as a prolate spheroid with semi-major axis $R_1$ along the magnetic field and semi-minor axis $R_2$ perpendicular. The mass-to-flux ratio is then
\begin{equation}
\frac{M}{\Phi_B}\bigg|_\mathrm{prolate} = \frac{4\pi R_1 {R_2}^2 \rho/3}{\pi {R_2}^2 B} = \frac{4}{3}\frac{R_1 \rho}{B}.
\end{equation}
The critical value for $R_1$ would be the same as given in Equation~(\ref{RcritSPH}), but the critical mass would be lower than that in Equation~(\ref{McritSPH}) by a factor $(R_2/R_1)^2$. For $R_1/R_2 \sim 3-4$, the critical mass will be similar to that found in the simulations.

Here we provide a physical picture for core formation via initial flow along the magnetic field, as illustrated in Figure~\ref{coreEvo} and \ref{Vtime}. Consider a post-shock layer with density $\rho_\mathrm{ps}$ and magnetic field $B_\mathrm{ps}$. For a cylinder with length $L$ along the magnetic field and radius $R$, the normalized mass-to-flux ratio is
\begin{equation}
\Gamma_\mathrm{cyl} = \frac{\pi R^2 L\rho_\mathrm{ps}}{\pi R^2 B_\mathrm{ps}}\cdot 2\pi\sqrt{G},
\end{equation}
and the critical length along the magnetic field for it to be supercritical is
\begin{equation}
L_\mathrm{mag, cyl} = \frac{B_\mathrm{ps}}{\rho_\mathrm{ps}}\frac{1}{2\pi\sqrt{G}}\label{theoLcrit}
\end{equation}
(note that up to a factor $3/4$, this is the same as Equation~(\ref{RcritSPH})).
The critical mass $M_\mathrm{mag,cyl}= \pi R^2 L_\mathrm{mag,cyl}\rho_\mathrm{ps} $ can then be written as 
\begin{equation}
M_\mathrm{mag,cyl} = \frac{R^2B_\mathrm{ps}}{2\sqrt{G}} = 1.2~\mathrm{M}_\odot \left(\frac{R}{0.05~\mathrm{pc}}\right)^2\left(\frac{B_\mathrm{ps}}{50~\mu\mathrm{G}}\right).
\label{Mcyl}
\end{equation}
This cylinder is gravitationally stable to transverse contraction unless $L\lesssim 2R$ \citep{1956MNRAS.116..503M}. However, contraction along the length of the cylinder is unimpeded by the magnetic field, and will be able to overcome pressure forces provided $L_\mathrm{mag,cyl}$ exceeds the thermal Jeans length, which is true in general for oblique shocks in typical conditions under consideration. The longitudinal contraction will produce an approximately isotropic core of radius $R$ when the density has increased by a factor
\begin{equation}
\frac{\rho'}{\rho_\mathrm{ps}} = \frac{L_\mathrm{mag,cyl}}{2R},
\label{rhoN}
\end{equation}
and at this point transverse contraction would no longer be magnetically impeded. For the core to have sufficient self-gravity to overcome thermal pressure at this point, the radius would have to be comparable to $R_\mathrm{th,sph}$ (see Equation~(\ref{RthSPH})):
\begin{equation}
R \sim R_\mathrm{th,sph} = 2.3\frac{c_s}{\sqrt{4\pi G \rho'}}.
\label{theoRBE}
\end{equation}
Combining Equations~(\ref{theoLcrit}), (\ref{rhoN}), and (\ref{theoRBE}) yields
\begin{equation}
\rho' = 0.19 \frac{{B_\mathrm{ps}}^2}{4\pi{c_s}^2},
\label{rho2}
\end{equation}
and
\begin{equation}
R = 5.3\frac{{c_s}^2}{\sqrt{G}B_\mathrm{ps}} = 0.05~\mathrm{pc} \left(\frac{B_\mathrm{ps}}{50~\mu\mathrm{G}}\right)^{-1}\left(\frac{T}{10~\mathrm{K}}\right).
\label{Rfinal}
\end{equation}
Substituting Equation~(\ref{Rfinal}) in Equation~(\ref{Mcyl}), the minimum mass that will be both magnetically and thermally supercritical, allowing for anisotropic condensation along $\mathbf{B}$, will be
\begin{equation}
M_\mathrm{crit} = 14 \frac{{c_s}^4}{G^{3/2}B_\mathrm{ps}} = 1.3\ \mathrm{M}_\odot \left(\frac{B_\mathrm{ps}}{50\mu\mathrm{G}}\right)^{-1}\left(\frac{T}{10~\mathrm{K}}\right)^2.\label{McritNew}
\end{equation}
Thus, anisotropic contraction can lead to low-mass supercritical cores, with values comparable to those formed in our simulations.\footnote{Note that up to factors of order unity, Equations~(\ref{rho2}) to (\ref{McritNew}) can equivalently be obtained by taking $B = B_\mathrm{ps}$ and requiring that the density $\rho\rightarrow\rho'$ in Equations~(\ref{MthSPH})$-$(\ref{RthSPH}) and (\ref{McritSPH})$-$(\ref{RcritSPH}) is such that $R_\mathrm{th,sph}\sim R_\mathrm{mag, sph}$ and $M_\mathrm{th,sph}\sim M_\mathrm{mag,sph}$.}

In addition, anisotropic condensation also helps explain why the core masses are quite similar for HD and MHD models, and independent of the angle between upstream magnetic field and converging flow. Note that Equation~(\ref{McritNew}) only depends on the post-shock magnetic field strength. For a magnetized shock, the post-shock magnetic pressure must balance the pre-shock momentum flux: ${B_\mathrm{ps}}^2/8\pi \sim \rho_0 {v_0}^2$. Therefore, Equation~(\ref{McritNew}) can be expressed as
\begin{equation}
M_\mathrm{crit} = 2.8 \frac{{c_s}^4}{\sqrt{G^3 \rho_0 {v_0}^2}} = 2.1~\mathrm{M}_\odot \left(\frac{n_0}{1000~\mathrm{cm}^{-3}}\right)^{-1/2} \left(\frac{v_0}{1~\mathrm{km/s}}\right)^{-1} \left(\frac{T}{10~\mathrm{K}}\right)^2.
\label{Mcritrhov}
\end{equation}
This is equivalent to Equation~(24) of \hyperlink{GO11}{GO11} with $\psi=2.8$. \hyperlink{GO11}{GO11} also pointed out that $\rho_0{v_0}^2$ will be proportional to $G{\Sigma_\mathrm{GMC}}^2$ for a gravitationally-bound turbulence-supported GMC. Thus, using Equation~(28) of \hyperlink{GO11}{GO11} in a cloud with virial parameter $\alpha_\mathrm{vir}$, Equation~(\ref{Mcritrhov}) would become
\begin{equation}
M_\mathrm{crit}=2.8~\mathrm{M}_\odot \left(\frac{T}{10~\mathrm{K}}\right)^2\left(\frac{\Sigma_\mathrm{GMC}}{100~\mathrm{M}_\odot~\mathrm{pc}^{-2}}\right)^{-1}{\alpha_\mathrm{vir}}^{-1/2}.
\label{Mcritalpha}
\end{equation}
Equations~(\ref{Mcritrhov}) and (\ref{Mcritalpha}) suggest that $M_\mathrm{crit}$ is not just independent of magnetic field direction upstream, it is also independent of magnetic field strength upstream. That is, when cores form in post-shock regions (assuming the GMC is magnetically supercritical at large scales), the critical mass is determined by the dynamical pressure in the cloud, independent of the cloud's magnetization. The models studied here all have the same dynamical pressure $\rho_0{v_0}^2$, and same upstream $B_0$. It will be very interesting to test whether for varying $B_0$ the core masses remain the same, and whether the scaling proposed in Equation~(\ref{Mcritrhov}) holds for varying total dynamic pressure.

\section{Summary}
\label{sec: summary}

In this work, we have used numerical simulations to study core formation in magnetized, highly dynamic environments, including the effect of ambipolar diffusion. Our simulations are fully three-dimensional, including a large-scale convergent flow, local turbulence, and self-gravity, and allow for varying ambipolar diffusion levels (parameterized by the ionization fraction coefficient $\chi_{i0}$) and shock obliquity (parameterized by the angle $\theta$ between the converging inflow and the global magnetic field). Filaments and then cores form in post-shock dense layers, with dense structures very similar to those found in observations. 

In all of our models (with or without ambipolar diffusion), magnetically supercritical cores form with physical properties similar to those found in observations. However, our parameter survey suggests that the transient ambipolar diffusion timescale and \textit{quasi-hydrodynamic} shocks are crucial in setting the magnetization of cores formed in post-shock regions. In addition, we demonstrate and quantitatively explain how low-mass supercritical cores form in strongly-magnetized regions, via anisotropic condensation along the magnetic field. 

Our main conclusions are as follows: 
\begin{enumerate}

\item
Under typical GMC conditions, isotropic formation of low-mass supercritical cores is forbidden under ideal MHD by the relatively strong magnetic support (Equation~(\ref{McritSPH})). This is true even downstream from strong MHD shocks where gas density is enhanced, because the magnetic field is compressed as well. In fact, for a spherical volume of given mass, the mass-to-flux ratio is generally larger for pre-shock conditions than post-shock conditions (Equation~(\ref{GammaComp}); except for the special case described in \#2 below). For typical conditions, the minimum post-shock critical mass for a spherical volume exceeds $10$~M$_\odot$ when ideal MHD applies (Table~\ref{modelpar}, \ref{simresults}). This suggests that either transient ambipolar diffusion in shocks must be taken into consideration, or that core formation is not spherically symmetric.

\item
When the incoming flows are almost parallel to the background magnetic field, MHD shocks will have compound post-shock conditions, including the regular \textit{fast} mode \citep{1992phas.book.....S} and the \textit{quasi-hydrodynamic} mode in which gas is compressed more strongly (Figure~\ref{rfcomp}). This happens when the angle $\theta$ between the inflow and the magnetic field is smaller than a critical value, $\theta_\mathrm{crit}$ (Equation~(\ref{thetaCrit})). For small $\theta$, the post-shock layer will have relatively high gas density and weak magnetic field compared to \textit{fast}-mode MHD shocks (Table~\ref{simresults}).

\item
Our three-dimensional simulations demonstrate the effect of transient ambipolar diffusion, as earlier identified and explained by \hyperlink{CO12}{CO12}. During the earliest stage of shock formation ($t\lesssim 0.3$~Myr), a thin but extremely dense layer appears in the middle of the shocked region in models with ambipolar diffusion (Figure~\ref{Evo1} and \ref{Evo2}), just like the central dense peak in the one-dimensional shocks analyzed by \hyperlink{CO12}{CO12}. Consequently, post-shock densities are generally higher in models with lower ionizations (smaller $\chi_{i0}$; see Table~\ref{simresults}), which correspond to stronger ambipolar diffusion as predicted in \hyperlink{CO12}{CO12}. 

\item
The ionization fraction is the main parameter controlling the transient ambipolar diffusion timescale needed for the gas to reach steady post-shock conditions ($t_\mathrm{transient}$). Models with smaller $\chi_{i0}$ have longer transient timescales (Equation~(\ref{t_tran})), indicating lower growth rate of the post-shock magnetic field and more weakly magnetized post-shock layers (Table~\ref{simresults}). Therefore, transient ambipolar diffusion is crucial in reducing the magnetic support in the post-shock regions (see $M_\mathrm{mag,sph}$ and $R_\mathrm{mag,sph}$ in Table~\ref{simresults}). 

\item
The filament network in more strongly magnetized post-shock cases is similar to those found in observations: in addition to large-scale main filaments, there are many thinner, less-prominent sub-filaments parallel to the magnetic field \citep{2008ApJ...680..428G,2011ApJ...734...63S,2012A&A...543L...3H,2013A&A...550A..38P,2013PPVI...Andre}. Dense cores form within the large-scale main filaments for all models.

\item
In our simulations, magnetically supercritical cores are able to form in the shock-compressed dense layers in all models, and the first collapse occurs at $t \lesssim 0.6$~Myr in most cases. Cores formed in our simulations have masses $\sim 0.04-2.5$~M$_\odot$ and sizes $\sim 0.015-0.07$~pc (Table~\ref{modelsum} and Figure~\ref{Dist}), similar to the values obtained in observations \citep[e.g.][]{2001A&A...372L..41M, 2009ApJ...691.1560I, 2009ApJ...699..742R,2013MNRAS.432.1424K}. The medians from the distributions are $0.47$~M$_\odot$ and $0.03$~pc. The mass-size relationship derived from our cores, $M\propto L^{2.3}$, also agrees with observations \citep[e.g.][]{1996ApJ...471..816E,2010MNRAS.402..603C,2010ApJ...723..492R,2013MNRAS.432.1424K}. 

\item
Our results show that the core mass and size are relatively independent of both the ambipolar diffusion and the upstream magnetic obliquity (Figure~\ref{MaModel}). Hydrodynamic and ideal MHD models also have very similar core masses and sizes. The core masses for ideal MHD cases with oblique shocks are more than an order of magnitude lower than the magnetic critical mass for a spherical region in the post-shock environment. Thus, simple estimates of the form in Equation~(\ref{McritSPH}) should not be used in predicting magnetically supercritical core masses from ambient environmental conditions in a GMC.

\item
The magnetic field of cores follows the same trends as the post-shock magnetization, in terms of variation with the upstream magnetic obliquity and ionization (Table~\ref{simresults}, \ref{modelsum}). This indicates that further ambipolar diffusion is limited during the core building phase, and instead cores form by anisotropic self-gravitating contraction as described in Section~\ref{sec::CF}. The mass-to-flux ratio in cores secularly increases with decreasing ionization (Figure~\ref{MaModel}), ranging from $\Gamma \sim 0.5$ to $7.5$ (Figure~\ref{GaDist}).
From all models combined, the median mass-to-flux ratio within cores is $\Gamma\sim 3$ (Figure~\ref{GaDist}), agreeing with the observed range of $\Gamma$ \citep[$\Gamma\sim 1-4$;][]{2008A&A...487..247F,2008ApJ...680..457T}. 

\item
Anisotropic self-gravitating condensation is likely the dominant mechanism for supercritical core formation in magnetized environments, regardless the magnetization strength and ionization fraction. Figures~\ref{coreEvo} and \ref{Vtime} clearly show how gas preferentially flows along the magnetic field lines in all models, creating dense cores that are both magnetically and thermally supercritical. The theoretical analysis of Section~\ref{anisotropic} shows that the characteristic mass expected from anisotropic contraction (Equation~(\ref{McritNew})) is similar to the median core mass obtained from our simulations (Figure~\ref{Dist}). For anisotropic core formation in a post-shock region, the critical mass is expected to depend only on the momentum flux entering the shock. We believe this explains why core masses in our simulations are similar regardless of the ionization level, whether the converging flow is nearly parallel to or highly oblique to the upstream magnetic field, or indeed whether the medium is even magnetized at all.

\end{enumerate}

\acknowledgements
We are grateful to Lee Mundy for many stimulating discussions, and to the referee for a helpful report. This work was supported by NNX10AF60G from NASA ATP, and by grant NNX13AO52H supporting C.-Y. C. under the NASA Earth and Space Science Fellowship Program.

\end{document}